\documentclass[12pt,letterpaper]{article}
\input epsf.tex
\usepackage{cite}
\usepackage{amsfonts}
\usepackage{amssymb}
\newcommand{\nc}{\newcommand}
\newcommand{\rnc}{\renewcommand}

\nc{\be}{\begin{equation}}
\nc{\ee}{\end{equation}}
\nc{\bea}{\begin{eqnarray}}
\nc{\eea}{\end{eqnarray}}
\nc{\ap}{\alpha^\prime}
\nc{\lam}{\lambda}
\nc{\dl}{\dot\lambda}
\nc{\ddl}{\ddot\lambda}
\nc{\dm}{\dot\mu}
\nc{\ddm}{\ddot\mu}
\nc{\dps}{\dot\psi}
\nc{\ddps}{\ddot\psi}
\nc{\ie}{{\it i.e.~}}
\nc{\eg}{{\it e.g.~}}
\nc{\trac}[2]{{\textstyle\frac{#1}{#2}}}

\nc{\ex}[1]{\mbox{e}^{\,\textstyle#1}}

\nc{\CC}{\Bbb{C}}
\nc{\HH}{\Bbb{H}}
\nc{\PP}{\Bbb{P}}
\nc{\RR}{\Bbb{R}}
\nc{\ZZ}{\Bbb{Z}}
\nc{\II}{\Bbb{I}}
\nc{\EE}{\Bbb{E}}

\rnc{\a}{\alpha}
\rnc{\b}{\beta}
\rnc{\d}{\delta}
\nc{\ga}{\gamma}
\nc{\la}{\lambda}
\nc{\f}{\phi}
\nc{\p}{\psi}
\nc{\e}{\eta}
\rnc{\c}{\chi}
\nc{\eps}{\epsilon}
\nc{\om}{\omega}
\nc{\Om}{\Omega}

\nc{\symx}{\circledS}
\nc{\ad}{\mathop{\mbox{ad}}\nolimits}
\nc{\tr}{\mathop{\mbox{tr}}\nolimits}
\nc{\Tr}{\mathop{\mbox{Tr}}\nolimits}
\nc{\Det}{\mathop{\mbox{Det}}\nolimits}
\rnc{\det}{\mathop{\mbox{det}}\nolimits}
\nc{\rk}{\mathop{\mbox{rk}}\nolimits}
\nc{\del}{\partial}
\nc{\diag}{\mathop{\mbox{diag}}\nolimits}
\nc{\ra}{\rightarrow}
\nc{\Ra}{\Rightarrow}
\nc{\LRa}{\Leftrightarrow}
\nc{\lra}{\leftrightarrow}
\nc{\ot}{\otimes}
\rnc{\ss}{\subset}
\nc{\nul}{\noindent\underline}
\nc{\non}{\nonumber\\}
\nc{\mat}[4]{\left(\begin{array}{cc}#1&#2\\#3&#4\end{array}\right)}
\rnc{\lg}{\frak{g}}
\nc{\G}[3]{\Gamma^{#1}_{\;{#2}{#3}}}
\nc{\nam}{\nabla_{\mu}}
\nc{\nan}{\nabla_{\nu}}
\nc{\dx}{\dot{x}}
\nc{\dxl}{\dot{x}^{\la}}
\nc{\dxm}{\dot{x}^{\mu}}
\nc{\dxn}{\dot{x}^{\nu}}
\nc{\ddx}{\ddot{x}}
\nc{\ddxm}{\ddot{x}^{\mu}}
\nc{\ddxn}{\ddot{x}^{\nu}}
\nc{\dxi}{\dot{\xi}}
\nc{\ddxi}{\ddot{\xi}}

\begin{document}
\begin{flushright} {\footnotesize IC/2006/088}  \end{flushright}
\vspace*{8mm}
%\vspace{0.5cm}

\def\thefootnote{\fnsymbol{footnote}}

\begin{center}

{\Large\bf The Effect of $\alpha^{\prime}$ Corrections in String Gas Cosmology}

\vspace{0.7cm}

{\large{Monica Borunda$^{\rm a}$ and  Lotfi Boubekeur$^{\rm b}$ }}

\vspace{1.2cm}

{\small\textit{${}^{\rm a}$Institut de Physique, University of Neuch\^atel,\\
Rue Breguet 1, CH-2000 Neuch\^atel, Switzerland.}}

\vspace{.3cm}

{\small\textit{${}^{\rm b}$Abdus Salam International Centre for Theoretical Physics \\
Strada Costiera 11, 34014 Trieste, Italy.}}

\vspace{.3cm}

\end{center}

\vspace{0.8cm}
\hrule \vspace{0.3cm} 
{\small  \noindent \textbf{Abstract} \\[0.3cm]
\noindent

In the Brandenberger-Vafa scenario of string gas cosmology, the Universe starts as a small torus of string length  dimension filled with a hot gas of strings. In such extreme conditions, in addition to the departure from Einstein gravity which is due to the dilaton, one expects higher curvature corrections to be relevant. Motivated by this fact, we study the effect of the leading $\alpha^\prime{}^3$ corrections in type IIB string theory for this scenario. Within the assumptions of: weak coupling, adiabatic evolution and thermodynamical equilibrium, we perturbatively solved the corresponding equations of motion in two different cases: (i) the isotropic case which is governed by a single scale factor and (ii) the anisotropic case given by two different scale factors. In the first case, we consider two regimes (ia) The Hagedorn regime where the string gas equation of state is that of  pressureless dust, and (ib) the radiation regime. In the second case, (ii), we only considered a radiation-like equation of state. We found that the inclusion of  $\alpha^\prime{}$ corrections affects the scale factor(s) in opposite way in the Hagedorn and in the radiation regimes, acting as a driving force for the first one and a damping force for the second one. This effect is small for reasonable initial conditions and it is only observed at early times. Morever it is bigger in the Hagedorn regime than in the radiation regime. We also analyzed the fixed points of the system. We found that there exists a stable dS fixed point, which does not exist when the corrections are neglected. 

\vspace{0.5cm}  \hrule

\def\thefootnote{\arabic{footnote}}
\setcounter{footnote}{0}

\newpage

\section{Introduction}
%%%%%%%%%%%%%%

String theory remains the most promising and theoretically appealing candidate for a complete and unified description of all Nature's forces including gravity. Given the energy scales where string theory is valid, cosmology offers an invaluable, and maybe the only, testing ground for  string theory ideas. String Gas Cosmology is one of the existing  approaches trying to obtain our observed 4D universe starting from a string theory set up. This approach was initiated by the pioneering work of Brandenberger and Vafa \cite{Brandenberger:1988aj}. It aims to address two important questions which are most relevant for cosmology: the initial cosmological singularity and the dimensionality of space-time. 

The initial singularity arises from the fact that physical quantities, such as the Ricci scalar and temperature, become singular as one extrapolates standard cosmology to very early times. String theory has a built-in symmetry called $T$-duality, which acts as $a(t)\to 1/a(t)$, where $a(t)$ is the scale factor, that prevents such a singularity. This means that a Universe with a tiny scale factor is equivalent to one with a very large scale factor. In this way, all the physical quantities remain well-behaved at all times. This feature is also used in the so-called pre big bang scenario \cite{Gasperini:2002bn} .

Another important issue is the dimensionality of the space-time. String theory (in its supersymmetric version) requires for consistency that the dimension of space-time is 10. If string theory is a valid description at early times, then what is the mechanism that made our 3 observed spatial dimensions grow and the remaining 6 become  unobservably small? In the Brandenberger and Vafa scenario all nine spatial dimensions start out toroidally compactified with a common radius of dimension of the string length $\ell_s\equiv\alpha'^{1/2}$, where $\alpha'$ is the Regge slope related to the string tension through $T=1/2\pi\ap{}$. The strings in this background will have both momentum (KK) and winding modes ($W$) whose dynamics will determine how many dimensions expand and how many stay at the string length. A crucial feature is that $W$-modes impede the expansion of the dimensions they are wrapping. The qualitative argument  is that it is easier for the $W$ and $\overline{W}$ to annihilate in the maximum dimension of space-time $D=4$ because of the  naive numerological argument 2+2=4.

Even though the Brandenberger-Vafa (BV) scenario is appealing, there exist obstacles to its concrete realization. The dynamics described above presupposes thermal equilibrium, which means that the annihilation rate of $W\overline{W}$ in the large dimensions has to be fast enough compared with the expansion of the Universe. Many studies have been done to see if this qualitative behavior survives, \eg in \cite{Sakellariadou:1995vk,Easther:2004sd}. According to \cite{Easther:2004sd} there are two possible situations: either all the dimensions remain small, or they expand and become large. In the first case,  the  $W\overline{W}$ annihilation is not efficient and a substantial number of these modes remain preventing the large dimensions to grow. The reason behind that is not the fact that the expansion of the Universe is very fast at the time of annihilation, but it is instead that the string coupling (which enters the $W\overline{W}$ annihilation cross section) becomes very small in a finite amount of time. In the second case, the universe starts with few $W$-modes, they annihilate completely and all dimensions decompactify. Only very special initial conditions lead to the desired outcome.

So far, many of  the existing studies of this scenario have been done using the low energy effective action of type IIB theory - dilaton gravity system neglecting $\alpha'$ corrections \cite{Easther:2004sd, Tseytlin:1991xk,Tseytlin:1991ss,  Alexander:2000xv,Brandenberger:2001kj,Easson:2001fy,Easther:2002mi,Watson:2002nx,
Easther:2002qk,Easther:2003dd,BBST,Borunda:2003xb,Boehm:2002bm,Alexander:2002gj,Campos:2003gj,Brandenberger:2003ge,
Watson:2003gf,Campos:2003ip,Patil:2004zp,Battefeld:2004xw,Berndsen:2004tj,Campos:2004yn,Battefeld:2005av,
Chatrabhuti:2006px,Kanno:2005ck,Kaya:2005qm,Patil:2005nm,Rador:2005vq,Rador:2005we,Rador:2005ib,Patil:2005fi,
Brandenberger:2005bd,Danos:2004jz}. However, since the dynamics are supposed to take place when the scale factors  are of order of the string length, it is then important to study the leading $\alpha^\prime$ corrections in the effective action. Due to the length scale we are dealing with, it is obvious that all  corrections will contribute in the same footing to the dynamics. In this paper, and as a modest step towards a full understanding of these corrections, we will address the effect of the leading   corrections in   type II string theory which are  proportional to $\ap{}^3 R^4$. 

This paper is organized as follows: In section \ref{sec:2} we summarize some aspects of the BV scenario in the absence of $\alpha^\prime$ corrections.  Then we summarize
the dynamics for two cases: a) the isotropic case, a single scale factor,  where we consider two different
regimes (equations of state), the Hagedorn and the radiation regimes; b) the anisotropic case, with two scale factors, where we consider the radiation regime. We analyse the dynamics of the system in the presence of the  leading $\alpha^\prime$ corrections, providing dilaton gravity equations for the isotropic case in section 3.1.
In sections 3.1.1 and 3.1.2, we find the perturbative solutions for the Hagedorn and the radiation regime.  The fixed points for the exact equations of motion and their stability are analysed in sections 3.1.3 and
3.1.4. For the anisotropic case, the perturbative solutions  are  found in section 3.2 and the presence of fixed points and  their stablity  are studied in sections 3.2.1 and 3.2.2. We present our conclusions in the
final section. In the Appendix, we collect some cumbersome formulae.

%%%%%%%%%%%%%%%%%%%%%%%%%%%%%%%%%%%%%%%%%%%%%%%%%%%%%%%%%

\section{The BV scenario without $\alpha^\prime$-corrections}

%%%%%%%%%%%%%%%%%%%%%%%%%%%%%%%%%%%%%%%%%%%%%%%%%%%%%%%%%%
%%%% 
\label{sec:2}

In the standard BV scenario the Universe starts out at a very high temperature, close to the Hagedorn temperature  $T_H=1/2\pi\sqrt{2 \alpha'}$, which is the limiting temperature in type IIB string theory \cite{Hagedorn:1965st}. 
The geometry of the Universe is taken as a 9 dimensional small torus described by the metric
\be
{\rm d}s^2= G_{AB}\,{\rm d}x^A\,{\rm d}x^B=-{\rm d}t^2 + \ell_s^2 \sum_{i=1}^9 a^2_i(t)\,{\rm d}\theta^2_i, \quad 0\le\theta_i\le2\pi \label{metric}
\ee
The universe is assumed to be homogeneous and in thermal equilibrium. The scale factors are parametrized as $a_i(t)= \ex{\lambda_i(t)}$. The torus radii are assumed to be all more or less equal to a common value given by the string length $\ell_s$. The Universe is assumed to be filled with a gas of closed strings\footnote{Even in the presence of $D$-branes of various dimensionalities, the dynamics is still dominated by fundamental strings  \cite{Alexander:2000xv}.}. For simplicity, the string gas is assumed to be an ideal gas of strings, whose free energy  $F(\lambda_i, \, \beta)$ at finite temperature $T=\beta^{-1}$ is computed from the string partition function $Z(\beta,\,\lambda_i)$ (see \eg \cite{BBST}) using the thermodynamical identity $F(\beta,\,\lambda_i)\equiv-\beta^{-1}\log Z(\beta,\,\lambda_i)$. For our purposes it is enough to include only massless string modes.  

To summarize, the standard BV scenario is based on two ingredients: the first ingredient, which comes from string theory, is the low energy effective action (at tree level in $\alpha'$ and without form fluxes) for type IIB string theory. The second ingredient, which comes from string thermodynamics, is the free energy  $F(\lambda_i, \, \beta)$ of the string gas. 

Thus, the action for the BV scenario is 
\be
S_{\rm{BV}}=S_0 + S_m, 
\ee
\begin{eqnarray}
S_0&=&{1\over 2\kappa^2_{10}}\int {\rm d}^{10}x\, \sqrt{-G}\,\ex{-2\phi}\left( R+ 4 (\del\phi)^2\right), \\
S_m&=&\int {\rm d}t\,\sqrt{-G_{00}}\, F(\lambda_i,\, \beta\sqrt{-G_{00}}),
\end{eqnarray} 
where $\phi=\phi(t)$ is the dilaton field and $\kappa_{10}$ is related to the string length through $\kappa^2_{10}={1\over2} (2\pi)^7 \alpha'^4$. 

The equations governing the dynamics are obtained by varying the action $S_{\rm{BV}}$ \cite{Tseytlin:1991xk}
\begin{eqnarray}
\dps^2-\sum_{i=1}^9\dl^2_i &=&\ex{\psi}\,E,\label{eq:sg1}\\
\ddl _i-\dl_i\dps &=&{1\over2}\ex{\psi}\,P_i,\label{eq:sg2}\\
\ddps-\sum_{i=1}^9\dl^2_i &=&{1\over2}\ex{\psi}\,E,\label{eq:sg3}
\end{eqnarray}
where the shifted dilaton is defined as $\psi\equiv 2\phi-{\displaystyle \sum_{i=1}^9}\lambda_i$. The pressure and the total energy of the string gas $P$ and $E$ are defined as 
\begin{eqnarray}
E&=&-2 \frac{\del S_m}{\del G_{00}},\\
P_i&=&-\frac{\del S_m}{\del \lambda_i}\label{pressure}.
\end{eqnarray} 
The conservation of the total energy reads
\be
\dot{E}+ \sum_{i=1}^9 \dl_i P_i=0.
\label{eq:consE}
\ee
Eq.~(\ref{eq:consE}) is equivalent to the adiabaticity condition which means that at any given temperature, the system will adjust its volume so to keep its entropy constant. Therefore, during the adiabatic evolution, there exists a one-to-one relationship $\{\lambda_i\}\leftrightarrow\beta^{-1}=T$.

It is worth mentioning that Eqs.~(\ref{eq:sg1}-\ref{eq:sg3}) are invariant under $T$-duality: 
\be 
\lambda_i \to-\lambda_i\;\; \textrm{and }\;\phi\to\phi -\sum_{i=1}^9 \lambda_i.\label{duality1}
\ee 
 Considering the string gas as a perfect gas is not only a simplifying assumption but it is also a very good approximation. Indeed, the string gas will have different equations of state depending on the temperature. For $T\simeq T_H$ (the Hagedorn regime), the string gas will behave as pressureless dust \ie  $P=0$, and the dimensions are  nearly frozen at their self-dual radius $\sim \sqrt{\alpha'}$. The general picture is that the universe starts out in a hot dense and isotropic phase close to the Hagedorn  temperature, then under the effect of some thermal fluctuations some dimensions will start expanding. The number of expanding dimensions is singled by the dynamics. As the temperature drops  below the Hagedorn temperature ($T< T_H$), the string gas will behave as radiation \ie $P={1\over 9}\,E$ (radiation regime).
In the following, we will consider two important subcases: the isotropic case where all the scale factors are the same, and the anisotropic case where 3 dimensions have already larger radii that the remaining 6.

\subsection{The isotropic case}

In this case, the background geometry is described by the metric (\ref{metric}), with all the scale factors equal  
\be
{\rm d}s^2 = -{\rm d}t^2+ \ell_s^2 \,\ex{2\lambda(t)} \sum_{i=1}^9{\rm d}\theta_i^2,  \quad 0\le\theta_i\le2\pi.
\ee

\subsubsection{Hagedorn regime}

As we said before, at temperatures close to $T_H$, the system will behave as pressureless dust. Furthermore, during this phase, the free energy vanishes  so the strings will have a constant energy given by
\be
E=\frac{S}{\beta_H}\equiv E_0\,,
\label{eqstatehagedorn}
\ee
where $S$ is the entropy. In this regime, and considering the case where all the scale factors are equal, one can find an approximate analytic solution for the system of Eqs.~(\ref{eq:sg1}-\ref{eq:sg3}) which is given by \cite{Tseytlin:1991xk,BBST}
\begin{eqnarray}
\ex{-\psi}&\simeq&\frac{E_0}{4}t^2 + Bt + \frac{B^2-dA^2}{E_0},\label{psianalytic}\\
\lambda&\simeq&\lambda_0+\frac{1}{\sqrt{d}}\log\left|\frac{(E_0t + 2B-2\sqrt{d}A)(B+\sqrt{d}A)}
{(E_0t + 2B+2\sqrt{d}A)(B-\sqrt{d}A)} \right|. \label{lanalytic}
\end{eqnarray}
$A$ and $B$ are integration constants related to the initial conditions and are given by
\begin{eqnarray}
A=\dl_0\,\ex{-\psi_0}, \quad B=-\dps_0\,\ex{-\psi_0}, 
\end{eqnarray}
and $d$ is the number of space dimensions ($d=9$). This analytic solution is in very good agreement with the solutions obtained numerically and is very useful to predict the asymptotic evolution as $t\to\infty$. From Eqs.~(\ref{psianalytic}) and (\ref{lanalytic}), we see that $a(t)$  tends to the asymptotic value  \cite{BBST}
\be
a_\infty=\ex{\lambda_0}\left| \frac{B+\sqrt{d}}{B-\sqrt{d}}\right|.
\ee
On the other hand, the shifted dilaton, and thus the dilaton, rolls monotonically to the weak
coupling regime.

\subsubsection{Radiation regime}

As the temperature decreases below $T_H$, the Universe enters the  radiation regime. The free energy of the string gas receives two main contributions corresponding to: {\sl (i)} the string massless modes, which we refer 
to as radiation contribution, and {\sl (ii)} the string massive modes which we refer to as matter contribution. In this paper, we will only consider massless modes since as it was seen in \cite{BBST} they provide the main contribution. The total energy reads \cite{BBST}
\begin{equation}
E^{(d)}_{rad}={d\; a^d\over 2 \pi} D(0)^2 \Gamma\left(d+1 \over 2 \right)\, (4\pi)^{d+1\over 2}\zeta(d+1)(1-2^{-(d+1)})\beta^{-d-1}, 
\label{energyradiation}
\end{equation}
where $D(N)$ is the degeneracy factor ($D(0)=16$) and $\zeta(d+1)$ is the Riemannian zeta-function.

The radiation contribution is characterized by the fact that the $d$  dimensions  exhibit a radiation-like equation of state, $P_d^{rad}=E^{rad}/d$. 
 By demanding that the universe undergoes an adiabatic evolution,
\be
\frac{\rm d}{{\rm d}t}S=0,
\ee
it follows that 
\be 
\beta=\beta_0\frac{a}{a_0}\, .
\ee
One expects that at sufficiently late times the dilaton has been already stabilized at weak coupling, this means, as noticed first in \cite{Tseytlin:1991xk}, that Eqs.~(\ref{eq:sg1}-\ref{eq:sg3}) admit only solutions with a radiation type equation of state. 

Introducing the two variables $x \equiv \dot{\lambda}$ and $y \equiv \dot{\psi}$ in the case of $d=9$, we can write 
Eqs.~(\ref{eq:sg1})-(\ref{eq:sg3}) as
\bea
\label{dotx}
\dot{x} &=& xy+\frac{1}{2} w(y^2-9x^2)\,, \\
\label{doty}
\dot{y} &=& \frac92 x^2+\frac12 y^2\,.
\eea
The fixed point of the system corresponds to $\dot{x}=0$ and 
$\dot{y}=0$. As shown in Ref.~\cite{Tseytlin:1991xk} the attractor
solutions satisfy the relation $x=-wy$.
Substituting this relation in  Eq.~(\ref{dotx}), we obtain 
the differential equation $\dot{x}=-(1+9w^2)x^2/2w$.
Integrating this equation gives the following asymptotic solution:
\bea
\label{withoutalpha}
\label{scale}
\dot{\lambda}=\frac{2w}{(1+9w^2)t}\,, \quad
a \propto t^{\frac{2w}{1+9w^2}}\,, \quad
\dot{\psi}=-\frac{2}{(1+9w^2)t}\,.
\eea

In the case of radiation ($w=1/9$) one has $a \propto t^{1/5}$. 
Since the time-derivative of the dilaton $\phi$ is given by 
$\dot{\phi}=(\dot{\psi}+9\dot{\lambda})/2$, the dilaton is 
stabilized for $w=1/9$, after the system enters the attractor 
stage characterized by $\dot{\lambda}=-w\dot{\psi}$. Note that 
if we consider the $d=3$ spatial dimensions with $w=1/3$
we recover the evolution of the scale factor in a radiation dominated  
universe \cite{BBST} ($a \propto t^{1/2}$).

Since the condition, $2w<1+9w^2$, is always satisfied independently 
of the equation of state, the universe does not exhibit an 
accelerated expansion in the absence of $\alpha'$ corrections.
Equation (\ref{withoutalpha}) also shows that both $\dot{\lambda}$
and $\dot{\psi}$ decreases toward zero with time.

\subsection{The anisotropic case (3+6)}

The background geometry in the anisotropic case takes the form

\be
{\rm d}s^2= -{\rm d}t^2 + \ell_s^2\sum_{i=1}^3 a^2_i(t)\,{\rm d}\theta^2_i + \ell_s^2\sum_{i=1}^6 b^2_i(t)\,{\rm d}\vartheta^2_i, \quad 0\le\theta_i\le2\pi \;\textrm{ and } \;0\le\vartheta_i\le2\pi,\label{metricanisotropic}
\ee
where $a_i(t)=a(t)=\ex{\lambda(t)}$ is the scale factor of the large dimensions
and  $b_i(t)=b(t)=\ex{\mu(t)}$ is the scale factor of the small dimensions.
Introducing three variables $x \equiv \dot{\lambda}$, 
$y \equiv \dot{\psi}$ and $z \equiv \dot{\mu}$, 
we find that Eqs.~(\ref{eq:sg1}-\ref{eq:sg3}) with an equation of state
\bea
\label{EOS2}
P_{\lambda}=\frac{E}{3}\,,~~~~
P_{\mu}=0\, ,
\eea
where $P_{\mu}$ ($P_{\lambda}$) is the pressure along the small (large) dimensions given by Eq.~(\ref{pressure}),  give
\bea
\label{dotxs}
& & \dot{x}=xy-\frac12 x^2 +\frac16 y^2 -z^2\,, \\
& & \dot{y}=\frac32 x^2+\frac12 y^2+3z^2\,, \\
& & \dot{z}=yz\,.
\eea
In Ref.~\cite{BBST} it was shown that the attractor
solutions satisfy
\bea
\label{att2}
y=-3x\,, \quad z=0\,.
\eea
As in the isotropic case, the time-derivative of the dilaton, 
$\dot{\phi}=(y+3x+6z)/2$, vanishes along the attractor solution.
Hence the dilaton is fixed in this regime.

Substituting Eq.~(\ref{att2}) in Eq.~(\ref{dotxs}), the variable 
$x$ satisfies the differential equation $\dot{x}=2x^2$.
Then in the attractor regime we obtain 
\bea
\label{att}
\dot{\lambda}=\frac{1}{2t}\,,\quad \dot{\mu}=0\,, \quad
a \propto t^{1/2}\,, \quad b={\rm const}\,,\quad
\dot{\psi}=-\frac{3}{2t}\,.
\eea
Hence the late-time evolution for the large dimensions is described 
by a radiation dominated phase with a fixed dilaton.
Since the pressure along the small dimensions vanishes, this leads to 
the attractor with $\mu={\rm const}$.
Hence the large dimensions expand due to the presence of radiation
whereas the small dimensions can be kept small relative to the large 
dimensions, provided that the initial conditions satisfy 
$\dot{\lambda}_0 > \dot{\mu}_0$ and $a_0 > b_0$.

 If one includes the massive modes they contribute with a pressure for the  large and small dimensions. The effect is such that the winding modes lead to a positive (negative) contribution to the large (small) dimensions and the momentum modes give a positive contribution to both large and small dimensions. However at the self-dual (initial) radius the pressure due to the winding modes in the small dimensions compensates the pressure due to the momentum modes resulting in a stabilization of the small dimensions.  On the other hand the large dimensions keep expanding.

%%%%%%%%%%%%%%%%%%%%%%%%%%%%%%%%%%%%%%%%%%%%%%%%
\section{The BV scenario including $\alpha'$ corrections}
%%%%%%%%%%%%%%%%%%%%%%%%%%%%%%%%%%%%%%%%%%%%%%%%
\label{sec:3}

Before including higher curvature corrections and studying their effects, let us discuss which 
\mbox{parame}-trization of the $\alpha^{\prime}$ correction will be used. 
It is well known that the low-energy dynamics of the massless string
modes can be described by an effective action. This action 
is defined as a field-theory action generating the S-matrix (scattering amplitudes) which
coincides with the massless sector of the (tree-level) string S-matrix \cite{Scherk:1971xy,Scherk:1974ca,Yoneya:1974jg,Scherk:1974mc}.  Since the perturbative S-matrix is
invariant under local field redefinitions of the background fields there exists a large class of effective
actions, which all correspond to the same string S-matrix \cite{Tseytlin:1986ti, Gross:1986iv}.
The particular choice of field variables corresponds to a particular choice of renormalization-group
scheme, for example, ``the $\sigma$-parametrization'' is the one most suitable for
comparison with the $\sigma$-model and the ``s-parametrization'' is more appropriate for
comparison with the string S-matrix. 
Such redefinitions affect the higher order in $\alpha^\prime$  terms of the target-space effective
action. However the equivalence theorem states that there exists a choice of ``relative'' scheme (\ie of  field redefinition) such that the perturbative solutions of the effective actions coincide \cite{Lovelace:1983yv,Candelas:1985en, Sen:1985qt,Callan:1985ia,Metsaev:1987zx}. Therefore for  simplicity we will choose to work in the $\sigma$-parametrization scheme.

We consider the action of type IIB strings in a ($d$+1)-dimensional 
space-time  including the leading $\alpha'$ corrections,  which in the $\sigma$-parametrization is
\be
\label{S}
S= S_0 +S_{\alpha'} + S_m\,,
\ee
where $S_{\alpha'}$ is given by \cite{Gross:1986iv, Grisaru:1986vi}
\be
\label{S0}
S_{\alpha'}={1\over2 \kappa^2_{d+1}}\int {\rm d}^{d+1}x\sqrt{-G}\ex{-2\phi}\left [\frac
18 \ap{}^3\zeta (3)
R_{\mu\nu\rho\sigma}R^{\alpha\nu\rho\beta}
(R^{\mu\gamma}_{\delta\beta}R_{\alpha\gamma}^{\delta\sigma}
-2R^{\mu\gamma}_{\delta\alpha}R_{\beta\gamma}^{\delta\sigma})
\right]\,.
\ee
In type I  and  heterotic string theory, the leading higher curvature correction is of $O(\alpha')$ and it is of the Gauss-Bonnet type. In type IIB, it turns out that, due to the higher number of supersymmetries, the leading higher curvature correction oppears only at $O(\alpha'^3)$. 

Our aim is to find solutions to the equations governing the dynamics including higher curvature corrections and to compare them with the corresponding solutions without the corrections. We would like to see how different the solutions are and when the corrections begin to be most relevant. These corrections are considered in all cases as perturbations to the original equations and are expected to be relevant only at early times where curvature can be large. If this were not the case, one would have to include next-to-leading corrections, which is an infinite series in powers of $\alpha'$. Thus, if next-to-leading terms are relevant this means that curvature is so large that the supergravity description will break down.

For our purposes it is important to recall the following two facts. First, since the equations including higher curvature corrections are of higher order with respect to  the case without the corrections and very complicated, it is very difficult to find analytic solutions and thus we will rely on numerical integration, which  gave excellent results in the standard case. Second, the corrections contain higher derivatives and higher powers of derivatives not present in the original equations. This fact is important for the structure of the solutions. Let $\Psi^{(0)}(t)\equiv\{ \lambda_i^{(0)},\, \psi^{(0)}\}(t)$ be the solutions of the equations without higher order corrections and let $\Psi^{(c)}(t)\equiv\{\lambda^{(c)}_i,\,  \psi^{(c)}\}(t)$ be the solution including the corrections.  Due to the above, the latter do not reduce to the ones without perturbations when taking the limit $\alpha'\to 0$, \ie they are not perturbative solutions. This fact makes  harder for us to find solutions that are physically meaningful. In order to find such solutions, we consider solutions that are perturbations of the original ones. We call these solutions perturbative solutions. 

Now, let us outline our strategy for finding the perturbative solutions. Given the above discussion about the properties of the equations, we will search for solutions that are perturbations over the original ones. Namely we will look for solutions of the form 
\begin{equation}
\Psi^{(c)}(t)=\Psi^{(0)}(t) + \alpha'^3\, \Psi^{(p)}(t),
\label{ansatz}
\end{equation}
where $\Psi^{(p)}(t)$ is the perturbation of the solution $\Psi^{(0)}(t)$. The consistency of the  ansatz Eq.~(\ref{ansatz}) implies that $\alpha'^3\, \Psi^{(p)}(t)\ll\Psi^{(0)}(t)$. Moreover, the ansatz (\ref{ansatz}) guarantees that the solution found will reduce to the non-perturbed one as $\alpha'\to0$. Plugging  this ansatz into the equations and expanding all the terms in powers of $\alpha'$, we obtain differential equations that determine $\Psi^{(p)}(t)$ knowing $\Psi^{(0)}(t)$. Then, we solve these equations by  imposing boundary conditions such  that the effect of the corrections at late time is negligible \ie  $\Psi^{(p)}(t_f)=0$, where $t_f$ stands for some late time. This corresponds physically to impose that at late times the asymptotic behavior of the solution is the same as the unpertubed one: at late times the dynamics must be
governed by the $R$ term.

\subsection{The isotropic case}

Let us first study the  situation in which the radii of the 9 dimensions
are all equal. For the action (\ref{S}) we obtain 
the following equations of motion: 
\bea
-9\dot\lambda^2+\dot\psi^2+ \alpha'^3 f_1&=&\ex{\psi} E\,, \label{eqmotcor1}\\
\ddl -\dl\dps+ \alpha'^3 f_2&=&\frac12  \ex{\psi} P\,,\label{eqmotcor2} \\
\ddps-9\dl^2+ \alpha'^3f_3&=&\frac12 \ex{\psi} E\,, \label{eqmotcor3}
\eea
where the correction terms are given by 
\bea
f_{1}&=&\zeta (3)
[1890\dl^8+216\dl^7\dps +396 \dl^5 \dps\ddl
-594\dl^4\ddl^2+216\dl^3\dps\ddl^2 -144\dl^2\ddl^3\nonumber\\
&&+36\dl\dps\ddl^3+27\ddl^4-396\dl^5\lambda^{(3)}
-432\dl^3\ddl\lambda^{(3)}-108\dl\ddl^2\lambda^{(3)}]\,, \label{f1}\\
f_{2}&=& 2\zeta (3) [60\dl^7\dps -36 \dl^5\dps\ddl
+6\dl^6(\dps^2-70\ddl -\ddps )
+22\dl^3\ddl (-3\dps\ddl +4\lambda^{(3)}) \nonumber\\
& &+4\dl\ddl^2(-5\dps\ddl+12\lambda^{(3)}) +11\dl^4(\dps^2\ddl
-\ddl\ddps -2\dps\lambda^{(3)}+\lambda^{(4)})  \nonumber\\
& &+6\dl^2(\dps^2\ddl^2+11\ddl^3-\ddl^2\ddps -4\dps\ddl
\lambda^{(3)}+2\lambda^{(3)}{}^2+2\ddl\lambda^{(4)})\nonumber\\
& &+\ddl (\dps^2\ddl^2 +8\ddl^3 -\ddl^2\ddps -6\dps\ddl\lambda^{(3)}
+6\lambda^{(3)}{}^2 +3\ddl\lambda^{(4)})]\,,\label{f2}\\
f_{3} &=& 18 \zeta (3)  [60 \dl^8+6 \dl^7 \dps
+6 \dl^6 \ddl-11 \dl^4 \ddl^2-2 \dl^2 \ddl^3+\ddl^4\nonumber\\
& &+\dl \ddl^2 (\dps \ddl-3 \lambda^{(3)})+
6 \dl^3 \ddl (\dps \ddl -2 \lambda^{(3)})+11 \dl^5 (\dps \ddl-\lambda^{(3)})] \label{f3}
\eea
 In the same way as Eqs.~(\ref{eq:sg1}-\ref{eq:sg3}) are invariant under $T$-duality,  Eqs.~  (\ref{eqmotcor1}-\ref{eqmotcor3}) should be invariant under the same symmetry. However the duality transformations Eq.~(\ref{duality1}) are modified as
\be 
\lambda_i \to-\lambda_i+\alpha^\prime{}^3f(\lambda_i , \phi )\;\; \textrm{and }\;\phi\to\phi -\sum_{i=1}^9 \lambda_i +\alpha^\prime{}^3g(\lambda_i , \phi )\, ,\label{duality2}
\ee
 where $f(\lambda_i , \phi )$ and  $g(\lambda_i , \phi )$ are complicated functions  which we will not calculate\footnote{See \eg \cite{Exirifard:2004ey,Exirifard:2005hh} for an explicit calculation of the duality transformations for particular metrics.} .
 
Plugging   the ansatz (\ref{ansatz}) into the equations of motion Eqs.~(\ref{eq:sg1}-\ref{eq:sg3}), we obtain differential equations for $\Psi^{(p)}(t)$, that are valid up to $O(\alpha'^3)$. They are given by

\bea
-18\dot\lambda^{(0)}\dot\lambda^{(p)}+2\dot\psi^{(0)}\dot\psi^{(p)}+f_1(\lambda^{(0)} ,\psi^{(0)})&=&\ex{\psi^{(0)}} E \psi^{(p)}\,, \label{eqmotcor11}\\
\ddl^{(p)} -(\dl^{(0)}\dps^{(p)}+\dl^{(p)}\dps^{(0)})+f_2(\lambda^{(0)} ,\psi^{(0)})&=&\frac12  \ex{\psi^{(0)}} P \psi^{(p)}\,,\label{eqmotcor22} \\
\ddps^{(p)}-18 \dl^{(0)}\dl^{(p)}+f_3(\lambda^{(0)} ,\psi^{(0)})&=&\frac12 \ex{\psi^{(0)}} E \psi^{(p)}\,, \label{eqmotcor33}
\eea
 We then integrate numerically these equations, using as an input both the analytic expression and the numerical solution for $\Psi^{(0)}(t)$ when available.

\subsubsection{Hagedorn regime}

As explained in section 2.1.1, during this phase the energy is nearly constant and the pressure is zero.
We adopted initial conditions around $E_0\sim 1000 \, \alpha^\prime{}^{-1/2}$, which corresponds to a high energy-Hagedorn regime, $a_0\sim 1$  and we have chosen the shifted dilaton to satisfy the condition $\ex{\psi} \ll1$ to ensure that the string coupling constant is initially small and hence the perturbation theory and the ideal gas approximation are trustable. We have carried out simulations for a wide variety of  initial values. Given $\lambda^{(0)}(0)$ and $\dot\lambda^{(0)}(0)$ the value of $\dot\psi^{(0)}(0)$ is fixed by the Hamiltonian constraint  (\ref{eq:sg1}).

Figure~\ref{figura1a}  shows the  evolution of the scale factor for the unperturbed solution $a^{(0)}(t)$ and the corrected solution $a^{(c)}(t)$.  At late times the effect of the corrections is negligible and the evolution is perfectly described by the unperturbed solutions Eqs.~(\ref{psianalytic}-\ref{lanalytic}). As one approaches smaller times,  the perturbed solution starts growing until it dominates completely  the unperturbed one. The higher curvature terms make the scale factor
become larger than in the case without corrections, thus causing the
expansion rate to become smaller. This is due to the fact that  $f_2(\lambda^{(0)},\psi^{(0)})$ in Eq.~(\ref{eqmotcor22}) acts as a negative potential at early times, vanishing at late times. 
This can be understood from Eq.~(\ref{f2}): a numerical evaluation shows that at small times the dominant contribution to $f_2(\lambda^{(0)},\psi^{(0)})$ is given by the
the third and fourth derivatives of $\lambda$, resulting in a negative contribution. Under the influence of this term, the expansion rate $\dot\lambda^{(c)}$ grows, while in the case without the corrections $\dot\lambda^{(0)}$ decreases, until it reaches the point where the corrections are not relevant anymore. At this point $\ddot\lambda^{(c)}$ changes sign and $\dot\lambda^{(c)}$ becomes indistinguishable from $\dot\lambda^{(0)}$.
Therefore the $\alpha^\prime$ corrections act as a driving force for the scale factor.  On the other hand, for negative  initial value of $\dot\lambda_0$, the effect is the opposite; the scale factor is smaller and the expansion rate is bigger, because now $f_2(\lambda^{(0)},\psi^{(0)})$ acts as a positive potential (damping force). 

Concerning the dilaton, in the uncorrected case it starts from its initial value at
$\phi^{(0)}=\phi_0$, in the case of Fig.~\ref{figura1a}  $\phi_0\simeq-2.5$, and decreases to more negative values.  However, when we include the corrections the corrected dilaton gets a positive contribution at early times, becoming less negative, thus increasing the string coupling, but remaining always in the weak coupling regime.

It is worth to notice that the  effect of the corrections might help to make the $W\overline{W}$
modes annihilate more efficiently. Indeed, including  the corrections
leads to both a decrease of the expansion rate and to an increase of
dilaton, and thus the string coupling, which might help to circumvent the
"all or nothing" problem in \cite{Easther:2004sd}.

How close  to $t=0$ one can trust the perturbed solution? As one approaches the origin the perturbation becomes bigger and therefore our analysis is not valid anymore. Moreover the Hamiltonian constraint must always be satisfied, with the correction terms included as well. It turns out that this energy constraint is satisfied up to  the inflection point of $a^{(c)}(t)$. 

Also, the initial conditions play an important role since they determine the magnitude and the time of appearance of the effect. The larger the value of $\dot\lambda^{(0)}(0)$ the larger the correction will be and therefore the effect will be noticable at later times. This makes the next-to-leading corrections in the $\alpha^\prime$ expansion larger and one would have to take them into account. 

%%%%%%%%%%%%%%%%%%%%%%%%%%%%%%%%%%%

\begin{center}
\begin{figure}[t]
\hspace{-1.3cm}
\epsfxsize = 7.5in \epsffile{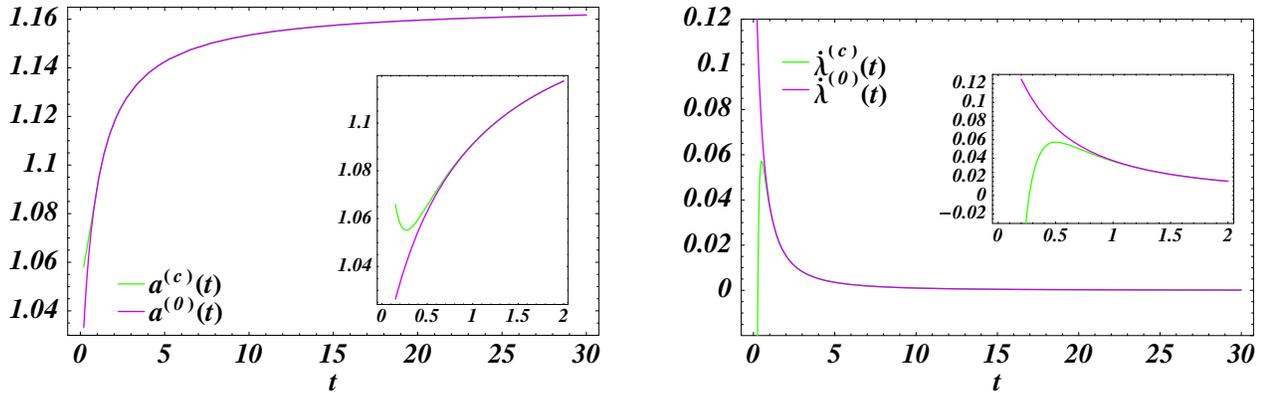} 
{\caption{ \label{figura1a}\small The evolution of the scale factor $a(t)$ and the expansion rate $\dot\lambda (t)$ for the initial conditions 
 $\lambda^{(0)}(0)=10^{-3}$, $\dot\lambda^{(0)}(0) =0.2$,  $\psi^{(0)}(0)=-5$ and $E_0=10^3$ in the Hagedorn
 regime.}}
\end{figure}
\end{center}
%%%%%%%%%%%%%%%%%%%%%%%%%%%%%%%%%

\subsubsection{Radiation regime}

In this case  we will consider  Eqs.~(\ref{eqmotcor11}-\ref{eqmotcor33}) with the equation of
state for radiation due to a gas of strings in a 9 dimensional space, $P=\frac 19 E$. In this
case the energy is given by Eq.~(\ref{energyradiation}) with $d=9$. 

Figure \ref{figura2}~ shows the behaviour for the corrected and the uncorrected  solutions.  As in the Hagedorn regime, the late time evolution is perfectly described by the uncorrected equations of motion Eqs.~ (\ref{eq:sg1}-\ref{eq:sg3}) and the effect of the corrected terms is visible only at very small values of time.

In contrast with the Hagedorn regime,  under the influence of $f_2(\lambda^{(0)},\psi^{(0)})$, the expansion rate decreases more rapidly, until the time where the corrections are not relevant anymore. There is no inflection point in this case. Once again, this can be understood by inspecting the sign of $f_2(\lambda^{(0)},\psi^{(0)})$ in Eq.~(\ref{eqmotcor22}) which in the  radiation regime is positive. Indeed,  the expansion rate in the radiation regime is bigger than in the Hagedorn regime, and therefore in Eq.~(\ref{eqmotcor22}), not only the last terms, which involve $\lambda^{(4)}$ and $\lambda^{(3)}$ will contribute, but also the remaining terms, resulting in a positive contribution. For negative values of $\dot\lambda^{(0)}(0)$, the result is the same. 

The dilaton in the uncorrected case starts from an initial negative value, for instance for the initial conditions in Figure \ref{figura2} $\phi_0\simeq-9.15$,  then it grows until stabilizing at weak coupling $\phi\simeq-4$. The effect of the corrections at early times is to give a negative contribution thus decreasing the string coupling constant. 

To summarize, there are  two main  differences with respect to the Hagedorn case. First, the effect of the corrected terms is stronger in the Hagedorn case than in the  radiation case, \ie for small values of
$\dot\lambda_0$ the effect due to the corrections is noticable in the Hagedorn
regime whereas it is negligible in the  radiation regime. Second, in the Hagedorn regime
the effect of the corrections results in a driving force for the scale factor, while in the radiation regime, it leads to a damping force. Furthermore, the dilaton gets a positive contribution in the Hagedorn regime and a negative one during the radiation regime.

%%%%%%%%%%%%%%%%%%%%%%%%%%%%%%%%%%%
\begin{center}
\begin{figure}[t]
\hspace{-1.3cm}
\epsfxsize = 7.5in \epsffile{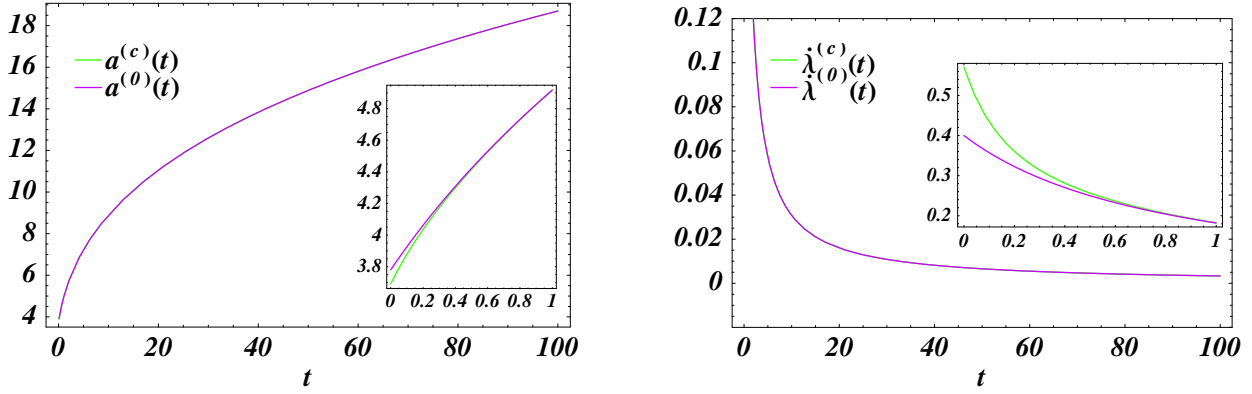} 
\caption{  \label{figura2} {\small The evolution of the scale factor $a(t)$   and the expansion rate $\dot\lambda (t)$ for the initial conditions 
$\lambda^{(0)}(0)=1.33$, $\dot\lambda^{(0)}(0) =0.4$,  $\psi^{(0)}(0)=-30$ and $\beta_0=15$ in the  radiation 
 regime.  }} 
\end{figure}
\end{center}
%%%%%%%%%%%%%%%%%%%%%%%%%%%%%%%%%%%

\subsubsection{Fixed points}
\label{fixed}

Even though the exact solutions to Eqs.~(\ref{eqmotcor1}-\ref{eqmotcor3}) are very difficult to obtain, one can look for the presence of fixed points.  Introducing new variables 
$\dot{\lambda}=y_1$, $\ddot{\lambda}=y_2$, 
$\lambda^{(3)}=y_{3}$ and $\dot{\psi}=y_{4}$, 
one can rewrite the Eqs.~(\ref{eqmotcor1}-\ref{eqmotcor3}) in an autonomous form:
\bea
& &\dot{y}_1=y_2\,, \\
& &\dot{y}_2=y_3\,, \\
\label{FG}
& &\dot{y}_3=G/F\,, \\
& &\dot{y}_4=\frac12 y_{4}^2+\frac92 y_{1}^2-
\ap{}^3 f_3+\frac12 \ap{}^3 f_1\,,
\eea
where 
\bea
F&=& 2\zeta(3) \ap{}^3 (11 y_1^4+
12y_1^2 y_2 +3y_2^2)\,, \\
G&=& -y_2+y_1 y_4+
\zeta(3) \ap{}^3 (6y_1^6+11y_1^4 y_2
+6y_1^2 y_2^2+y_2^3)(y_4^2+9y_1^2) \nonumber \\
& &+\frac{w}{2} (-9y_1^2+y_4^2+\ap{}^3f_1)
-\ap{}^3 f_4\,,
\eea
and $f_4$ is defined in the Appendix.

The fixed points of this system can be obtained by setting 
$\dot{y}_i=0$ $(i=1, 2, 3, 4)$ \cite{STTT,CTS}. This gives 
\bea
\label{fix1}
& &9y_1^2+y_4^2=270\zeta(3)\ap{}^3y_1^8\,, \\
\label{fix2}
& & y_1y_4+\frac{w}{2}(-9y_1^2+y_4^2)+
6\zeta(3)\ap{}^3 \left[-20y_1^7 y_4
-2y_1^6 y_4^2 +\frac{w}{2}
(315 y_1^8+36y_1^7 y_4)
\right]=0\,. \nonumber \\
\eea
Combining these two equations, we find
\bea
[y_1-3\zeta(3) \ap{}^3(40y_1^7 
+4y_1^6 y_4)]
(y_4-9w y_1)=0\,.
\eea
In the rest of this subsection we shall set $\ap{}^3=1$ \footnote{$\alpha^\prime{}^3$ can  be recovered by $\zeta (3)\to \ap{}^3\zeta (3)$.}.

When $y_1-3\zeta(3) (40y_1^7 
+4y_1^6 y_4)=0$, one obtains
\bea
\label{ia}
& &{\rm (ia)}\,\,\textrm{de  Sitter}:\,\,\dot{\lambda}=0.8282, \quad 
\dot{\psi}=-8.1043\,, \\
\label{ib}
& &{\rm (ib)}\,\, \textrm{anti de  Sitter}:\,\,
\dot{\lambda}=-0.8282, \quad 
\dot{\psi}=8.1043\,,
\eea
where we used $\zeta(3)=1.20206$.
These fixed points exist for any value of $w$.
The point (ia) is a de Sitter solution in which the scale factor
grows exponentially ($a \propto \ex{\dot\lambda t}$), 
whereas the point (ib) is an anti de Sitter solution in which 
$a$ decreases exponentially.
On the other hand, when $y_4=9w y_1$, we obtain the following fixed points:
\bea
\hspace*{-1.0em}& &{\rm (iia)}\,\, \textrm{de Sitter}:\,\, \dot{\lambda}=
\left(\frac{1+9w^2}{30\zeta(3)}\right)^{1/6}, \quad 
\dot{\psi}=9w\left(\frac{1+9w^2}{30\zeta(3)}\right)^{1/6}\,, \label{iia}\\
\hspace*{-1.0em}& &{\rm (iib)}\,\, \textrm{anti de Sitter}: \,\,
\dot{\lambda}=
-\left(\frac{1+9w^2}{30\zeta(3)}\right)^{1/6}, \quad 
\dot{\psi}=-9w\left(\frac{1+9w^2}{30\zeta(3)}\right)^{1/6}\label{iib}\,,
\eea
which are dependent on $w$. Solutions with the  fixed points (\ref{ia}-\ref{iib}) correspond
to non-perturbative in $\alpha^\prime$ solutions, \ie   $(\dot\lambda , \dot\psi )\sim\alpha^\prime{}^{-1/2}$.

On the other hand, let us  note that there also exists a perturbative Minkowski fixed point:
\bea
{\rm (iii)}\,\, {\rm Minkowski}: \,\,\dot{\lambda}=0\,, 
\quad \dot{\psi}=0\,.
\eea

In the following we shall study the stability of these points against 
perturbations and then clarify the attractor of the system.

\subsubsection{The stability of fixed points}

Consider small perturbations, $\delta y_{i}$, about the 
fixed points $y_{i}^{(f)}$ derived in the previous subsection, \ie  
$y_{i}=y_{i}^{(f)}+\delta y_{i}$. Using the fact that $y_2^{(f)}=y_3^{(f)}=0$ and $G^{(f)}=0$ 
at the fixed points, we obtain the following perturbation equations:
\bea
\label{dely1}
& & \delta \dot{y}_{1}=\delta y_{2}\,, \\
& & \delta \dot{y}_{2}=\delta y_{3}\,, \\
\label{delG}
& & \delta \dot{y}_{3}=\frac{1}{F} \delta G\,, \\
\label{dely4}
& & \delta \dot{y}_{4}=
9y_{1} \delta y_{1}+y_{4} \delta y_{4}-\delta f_{3}
+\frac12 \delta f_{1}\,.
\eea
Here $\delta f_{i}$ and $\delta G$ are given by 
\bea
&\delta f_{1}=\displaystyle{\sum_{i=1}^4} c_i \delta y_{i}\,, \quad
\delta f_{3}=\sum_{i=1}^4 d_i \delta y_{i}\,, \quad
\delta f_{4}=\sum_{i=1}^4 e_i \delta y_{i}\,,&\\
&\delta G=\displaystyle\sum_{i=1}^4 s_{i} \delta y_{i}\,,&
\eea
and $c_i$, $d_i$, $e_i$ and  $s_i$ are defined in the Appendix. 
Equation (\ref{dely4}) yields
\bea
\delta \dot{y}_{4}=(9y_{1}-1080\zeta(3)\ap{}^3 y_1^7)
\delta y_{1}-108 \zeta(3)\ap{}^3 y_1^6 \delta y_{2}
+y_{4} \delta y_{4}\,.
\eea

Hence the differential equations (\ref{dely1})-(\ref{dely4}) can be written 
in the form: 
\bea
{}^t (\dot{y}_{1}, \dot{y}_{2}, \dot{y}_{3}, 
\dot{y}_{4})={\cal M}\,\, {}^t(y_1, y_2, y_3, y_4)\, ,
\eea
with 
\be
{\cal M}=\left(\begin{array}{cccc}
0&1&0&0\\
0&0&1&0\\
s_1/F&s_2/F&s_3/F&s_4/F\\
9y_{1}-1080\zeta(3) y_1^7&-108 \zeta(3) y_1^6&0&y_4
\end{array}\right)\, .
\ee

If all eigenvalues $v$ of the matrix ${\cal M}$ are negative or have 
negative real parts, the fixed points are regarded to be stable.
Meanwhile if  at least one of the eigenvalues is positive, the fixed point
is not a stable attractor.

We numerically evaluate the eigenvalues for the equation of state
$0 \le w<1$ and find the following properties for the fixed points
(ia), (ib), (iia), (iib) and (iii) introduced in subsection \ref{fixed}.

\begin{itemize}

\item (ia) All eigenvalues are negative. 
When $w=1/9$, for example, one has 
\bea
v=-1.84, -6.26, -8.10, -8.9.
\eea
Hence this de Sitter fixed point is stable.

\item (ib) All eigenvalues are positive. 
When $w=1/9$, for example, one has 
\bea
v=1.84, 6.26, 8.10, 8.9.
\eea
Thus this anti de Sitter fixed point is an  unstable node.

\item (iia) Only one of the eigenvalues is negative and other 
eigenvalues are positive or have positive real parts.
When $w=1/9$, for example, one has 
\bea
v=0.28+1.00i, 0.28-1.00i, -3.34, 3.90.
\eea
Thus this de Sitter fixed point is unstable.

\item (iib) Three of the eigenvalues are negative or have 
negative real parts, but one of the eigenvalues is positive.
When $w=1/9$, for example, one has 
\bea
v=-0.28+1.00i, -0.28-1.00i, 3.34, -3.90.
\eea
Thus this anti de Sitter fixed point is unstable saddle.

\item (iii) The term $F$ in Eq.~(\ref{delG}) is given by 
$F=22\zeta(3)\ap{}^3  y_{1}^4$ and the term $s_{4}$
includes the $y_1$ term. This means that the $s_{4}/F$
term on the RHS of Eq.~(\ref{delG}) exhibits a divergence
for $y_{1} \to 0$.
Hence the Minkowski fixed point is unstable.

\end{itemize}    

{}These  results indicate that the de Sitter fixed point (ia) 
is the only stable attractor. 
This is confirmed by our numerical simulations shown 
in Figure \ref{phase}.
At this fixed point the time-derivative of the dilaton $\phi$
is negative, \ie  
$\dot{\phi}=(\dot{\psi}+9\dot{\lambda})/2=-0.3253$, 
which means that the system approaches a weak coupling 
regime. Hence the description of an ideal string gas
based on a canonical ensemble is valid in this attractor regime.

%%%%%%%%%%%%%%%%%%%%%%%%%%%%%%%%%%%%
\begin{center}
\begin{figure}[t]
\begin{center}
\epsfxsize = 3.5in \epsffile{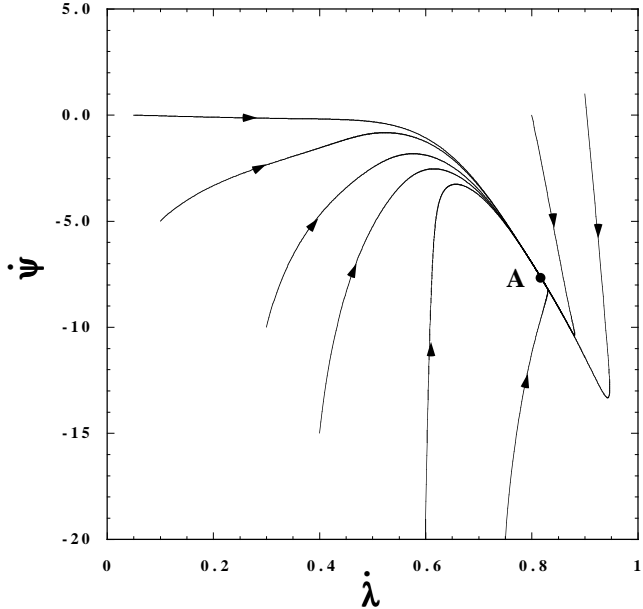} 
\caption{\label{phase}  \small Phase space plot in terms of $\dot{\lambda}$
and $\dot{\psi}$ for $w=1/9$ with initial conditions 
$\ddot{\lambda}=0.5$ and $\lambda^{(3)}=0$.
The solutions approach the 
de Sitter fixed point (ia) [$\dot{\lambda}=0.8282$
and $\dot{\psi}=-8.1043$], which is denoted as 
the point A in the figure.}
\end{center}
\end{figure}
\end{center}
%%%%%%%%%%%%%%%%%%%%%%%%%%%%%%%%%%%%

We recall that the effect of $\ap{}$ corrections is important when 
$\dot{\lambda}$ is order unity (note that we fixed $\ap{}=1$).
The de Sitter fixed point (ia) [$\dot{\lambda}=0.8282$] exists
in a high-curvature regime in which the $\ap{}$ corrections lead
to geometrical inflation.  It is interesting to note that the effect of $\ap{}^3$ 
corrections leads to an inflationary solution, while this can not 
be obtained without $\ap{}$ corrections.
This type of geometrical inflation often appears in the context of 
second-order string gravity \cite{CTS}.
{}From the above argument the effect of curvature corrections can be 
important even when the dilaton evolves toward the weakly coupled 
regime. This implies that ``the high-curvature regime'' is generally 
different from ``the strong coupling regime''. However one must be careful because at this value of $\dot\lambda$ other subleading
order $\alpha^\prime$ corrections may have to be included to perform a complete analysis.

In the absence of $\ap{}$ corrections the asymptotic behaviour
of the scale factor for the radiation like equation of state is given by Eq.~(\ref{scale})
with $\dot{\lambda}$ and $\dot{\psi}$ decreasing 
towards 0. Meanwhile we showed that the Minkowski fixed 
point $(\dot{\lambda}, \dot{\psi})=(0, 0)$ is unstable
in the presence of $\ap{}$ corrections.
In fact we have numerically confirmed this instability 
around the fixed point for initial conditions of 
$\dot{\lambda}, \ddot{\lambda}, \lambda^{(3)}, \dot{\psi}$ 
close to zero. The solutions tend to approach the de Sitter 
fixed point (ia) provided that $\ddot{\lambda}$ are not very 
small relative to 1.

%%%%%%%%%%%%%%%%%%%%%%%%
\subsection{The anisotropic case}
%%%%%%%%%%%%%%%%%%%%%%%%

Given the metric  (\ref{metricanisotropic}) the variation of the action (\ref{S0}) in the presence of an 
ideal string gas leads to the equations of motion:
\bea
\label{bes1}
-3\dot\lambda^2-6\dot{\mu}^2
+\dot\psi^2+ \ap{}^3 h_1&=&\ex{\psi} E\,, \\
\label{bes2}
\ddl -\dl\dps+\ap{}^3  h_2&=&\frac12  
\ex{\psi} P_{\lambda}  \,, \\
\label{bes3}
\ddot{\mu}-\dot{\mu} \dot{\psi}+\ap{}^3  h_3
&=&\frac12 \ex{\psi} P_{\mu}\,, \\
\label{bes4}
\ddps -3\dl^2-6\dot\mu^2+\ap{}^3 h_4&=&\frac12 \ex{\psi} E\,,
\eea
where $P_{\lambda}$ and $P_{\mu}$ are the pressure of the string gas along  the large  and small dimensions, respectively.
Explicit forms of the $h_{1}$, $h_{2}$, $h_{3}$ and $h_{4}$ terms are presented in the Appendix.

As discussed  in Sec. 2, the pressures $P_{\lambda}$ and $P_{\mu}$ in the radiation regime are
given by  Eq.~(\ref{EOS2}). We are interested in the evolution of the large and small dimensions in 
the presence of $\ap{}$ corrections. By plugging in the ansatz (\ref{ansatz}) into Eqs.~(\ref{bes1}-\ref{bes4})  
one obtains  a set of differential  equations analogous to  Eqs.~(\ref{eqmotcor11}-\ref{eqmotcor33}) for $\Psi^{(c)}(t)$. Therefore using the solutions for the uncorrected equations $\Psi^{(0)}(t)$ we can numerically solve for $\Psi^{(c)}(t)$.

As it was shown in \cite{BBST}, in the uncorrected case, \ie $h_i=0$, $i=1,\cdots,4$ the expansion
rate for the small dimensions $\dot\mu$ is exponentially suppressed with the decrease of $\psi$ since in this case  the pressure $P_\mu$ is zero and Eq.~(\ref{bes3}) integrates to  $\dot\mu =\dot\mu_0 \ex{\psi-\psi_0}$. Therefore in the absence of $\alpha^\prime$ corrections unless the initial value of $|\dot\mu |$ is much larger than unity, the radius $b$ can stay small around $b\sim 1$.  It was also shown that the matter contribution,
given by the massive string modes, is such that the pressure due to the winding and the momentum modes cancels  at values $b\sim 1$ leading to a stablization of the small dimensions. Hence we shall only consider initial values of $b\sim 1$. Moreover, solutions with growing $a$ and small nearly constant $b$ were found
for initial values $a_0>b_0\sim 1$, $\dot\psi <0$ and $\dot\lambda_0 \geq |\dot\mu_0|$. Figures \ref{figura3} and \ref{figura4} show such a solution for initial conditions $\dot\lambda^{(0)}(0)=0.6$ and
$\dot\mu^{(0)}(0) =0.1$.  The effect of the higher curvature terms is basically the same as in the
isotropic case with the same equation of state (radiation). In this case, $h_2$ and $h_3$ in  Eqs.~(\ref{bes1}-\ref{bes2}) result in a positive potential for $\dot\lambda$ and $\dot\mu$ at early times. Thus at initial times the expansion rate of both dimensions decreases more rapidly. However, the effect is less dramatic for $\dot\mu^{(c)}$ than for $\dot\lambda^{(c)}$. This is due to the fact that the large dimensions receive also a  
 contribution from the pressure in contrast with the small ones where there is no pressure. When $\dot\lambda \ll |\dot\mu|$ holds initially it was shown in \cite{BBST}  that in the absence of $\alpha^\prime$ corrections it is difficult to keep the small dimensions small relative to the large ones and the  effect of the corrected terms, at early times, is basically the same as in the previous case.  For negative values of $\dot\lambda_0$ we get bouncing solutions as in the case without corrections \cite{BBST}. The large dimensions start contracting, then $\dot\lambda^{(c)}$ changes sign and then they start to grow. The effect of the corrections on the dilaton is basically the same as in the isotropic case with radiation-like equation of state.

%%%%%%%%%%%%%%%%%%%%%%%%%%%%%%%%%%%
\begin{center}
\begin{figure}[t]
\hspace{-1.7cm}
\epsfxsize = 7.5in \epsffile{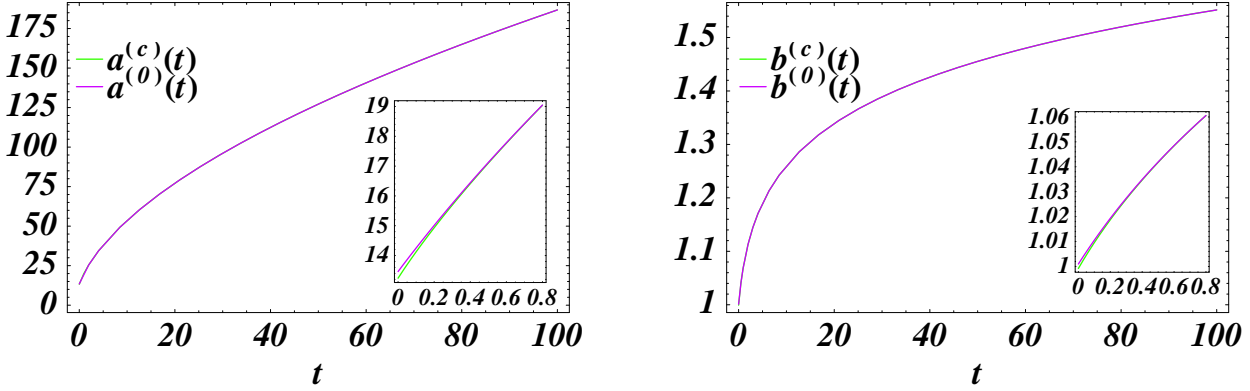} 
\caption{ \label{figura3}{\small The evolution of the scale factors $a(t)$   and $b(t)$  for the initial conditions 
$\lambda^{(0)}(0)=2.6$, $\dot\lambda^{(0)}(0) =0.6$, $\mu^{(0)}(0)=10^{-3}$, $\dot\mu^{(0)}(0)=0.1$,  $\psi^{(0)}(0)=-20$ and $\beta_0=15$ in the  radiation regime.  }} 
\end{figure}
\end{center}
%%%%%%%%%%%%%%%%%%%%%%%%%%%%%%%%%%%

\subsubsection{Fixed points}

In this subsection we shall derive the fixed points of the system of Eqs.~(\ref{bes1}-\ref{bes4}).  In analogy
 to the isotropic case the fixed points can be obtained by setting $\ddot{\lambda}=0$, $\lambda^{(3)}=0$, 
$\lambda^{(4)}=0$ and $\ddot{\psi}=0$ in Eqs.~(\ref{bes1})-(\ref{bes4}), which gives
\bea
-3\dl^2-6\dm^2+\dps^2+\ap{}^3 h_1(\ddot{\lambda}=\lambda^{(3)}=\lambda^{(4)}=\ddot{\psi}=0)
& =&\ex{\psi} E\,,\\
 -\dl\dps+\ap{}^3 h_2(\ddot{\lambda}=\lambda^{(3)}=\lambda^{(4)}=\ddot{\psi}=0)
&=&\frac 12 \ex{\psi} P_{\lambda}\,, \label{Plam}\\
-\dm\dps+\ap{}^3 h_3(\ddot{\lambda}=\lambda^{(3)}=\lambda^{(4)}=\ddot{\psi}=0)
&=&\frac 12 \ex{\psi} P_{\mu}\,,\label{Pmu}\\
{{\dps }^2}+\ap{}^3 h_1(\ddot{\lambda}=\lambda^{(3)}=\lambda^{(4)}=\ddot{\psi}=0)
-\ap{}^3  h_4(\ddot{\lambda}=\lambda^{(3)}=\lambda^{(4)}=\ddot{\psi}=0)
&=&\frac 12 \ex{\psi} E\,.\eea

Let us consider the equation of state of the string gas given by 
Eq.~(\ref{EOS2}).
Then we find from Eq.~(\ref{Pmu}) that 
$\dot{\mu}=0$ is the solution of Eq.~(\ref{Pmu}).
In this case we obtain the following fixed points:
\bea
\label{ideal}
(\dot{\lambda}, \dot{\mu}, \dot{\psi})&=&
(0.84719, 0, 0.84719), (-0.84719, 0, -0.84719), \nonumber \\
& &(1.0852, 0, -4.1566), (-1.0852, 0, 4.1566)\,.
\eea
The first and third points correspond to de Sitter solutions
for the large dimensions ($a \propto \ex{\dot{\lambda}t}$)
with the small dimensions being stabilized ($b={\rm const}$).
If we impose the validity of the weak coupling approximation,
the third one can be regarded as an ideal solution if it is 
a stable fixed point. However we shall see that this is not 
a stable attractor. 

Analogous to the fixed points (\ref{ia}) and (\ref{ib}) in the isotropic case,
there exist two fixed points in which both large and small dimensions
evolve with the same rate:
\bea
\label{fixedls}
(\dot{\lambda}, \dot{\mu}, \dot{\psi})=
(0.8282, 0.8282, -8.1043), (-0.8282, -0.8282, 8.1043)\,.
\eea
In fact by setting $\dot{\lambda}=\dot{\mu}$ in Eqs.~(\ref{Plam}) and 
(\ref{Pmu}), the RHS of these equations are the same.
Since $P_{\mu}=0$, we obtain $1-3\zeta(3)\ap{}^3
(40\dot{\lambda}^6+4\dot{\lambda}^5 \dot{\psi})=0$
for $\dot{\lambda} \dot{\psi} \neq 0$.
This is exactly the same equation as the one which we used 
to derive the fixed points (\ref{ia}) and (\ref{ib}), thereby 
giving the fixed points given above.
It is worth mentioning that  $P_{\lambda}$, $P_{\nu}$ and 
$E$ vanish in this case. This clearly shows that the fixed points
(\ref{fixedls}) exist irrespective of the presence of the string gas.
As in section 3.1.3 solutions with the above fixed points correspond
to non perturbative in $\alpha^\prime$ solutions.
On the other hand the  perturbative Minkowski fixed point 
$(\dot{\lambda}, \dot{\mu}, \dot{\psi})=(0, 0, 0)$ is present as well.

%%%%%%%%%%%%%%%%%%%%%%%%%%%%%%%%%%%
\begin{center}
\begin{figure}[t]
\hspace{-1.7cm}
\epsfxsize = 7.5in \epsffile{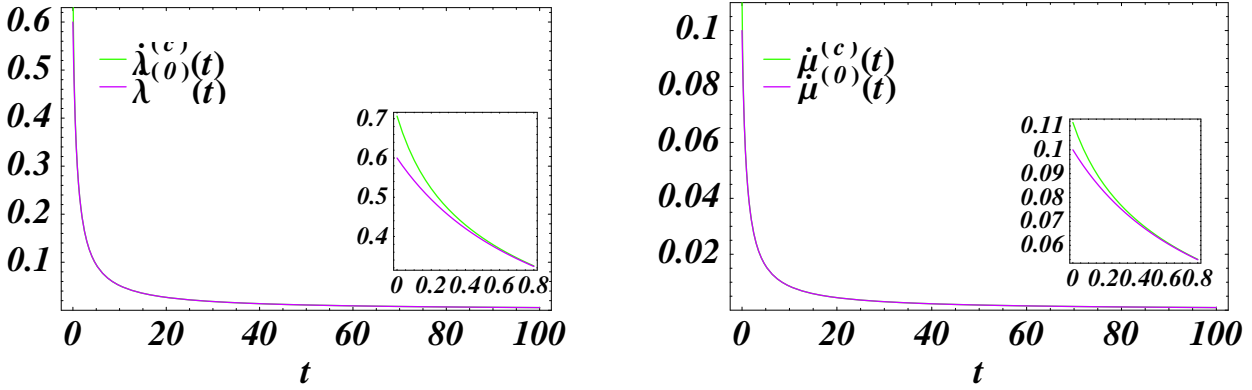} 
\caption{\label{figura4}{\small The evolution of the   expansion rates $\dot\lambda (t)$ and $\dot\mu (t)$ for the initial conditions  $\lambda^{(0)}(0)=2.6$, $\dot\lambda^{(0)}(0) =0.6$, $\mu^{(0)}(0)=10^{-3}$, $\dot\mu^{(0)}(0)=0.1$,  $\psi^{(0)}(0)=-20$ and $\beta_0=15$ in the  radiation regime.  }} 
\end{figure}
\end{center}
%%%%%%%%%%%%%%%%%%%%%%%%%%%%%%%%%%%

%
\subsubsection{Stability of the fixed points}
From the discussion in Sec. 3.1.4 it is clear that 
the  de Sitter fixed point in Eq.~(\ref{fixedls})
is stable whereas the  anti de Sitter one is unstable.
The Minkowski fixed point is not stable as in the isotropic case.
One can study the stability of fixed points (\ref{ideal})
by considering linear perturbations.
Because of the complexity of the $\ap{}$ corrections given in 
the Appendix,
we do  not consider perturbation equations in the anisotropic case.
Instead we shall numerically solve the evolution equations  
(\ref{bes1})-(\ref{bes4}) for a wide range of initial conditions
and see the stability of fixed points in the next subsection.
It turns out that the fixed points (\ref{ideal}) are unstable.

For numerical purpose it is convenient to write the equations in an 
autonomous form.
{}From Eqs.~(\ref{bes1}) and (\ref{bes4}) we get 
\bea
\label{ddotpsi}
\ddot{\psi}=\frac12 \dot{\psi}^2+\frac32 \dot{\lambda}^2+
3\dot{\mu}^2-\ap{}^3  h_4+\frac12 \ap{}^3  h_1\,.
\eea
Combining Eqs.~(\ref{bes2}) and (\ref{bes3}) with the use of 
Eq.~(\ref{ddotpsi}), we obtain
\bea
& & A\lambda^{(4)}+2B\mu^{(4)}=F\,, \\
& & B\lambda^{(4)}+C\mu^{(4)}=G\, ,
\eea
and therefore
\be
\label{lammu4}
\lambda^{(4)}=\frac{2BG-CF}{2B^2-AC}\,, \quad\quad
\mu^{(4)}=\frac{BF-AG}{2B^2-AC}\,,
\ee
where
\bea
A&=&\frac 12 \zeta (3)\ap{}^3[3(\ddl^2+\ddm^2)+(11\dl^4+3\dm^4)+6(\dm^2\ddm+2\dl^2\ddl)
+3\dl\dm (\dl\dm+2\dm^2+2\ddm)]\, ,\nonumber\\
B&=&\frac 14 \zeta (3)\ap{}^3[3\dl\dm(2(\dl^2+\dm^2+\ddl+\ddm)+3\dl\dm)
+6(\dm^2\ddl+\dl^2\ddm+\ddl\ddm)]\, ,\nonumber\\
C&=&\frac 14 \zeta (3)\ap{}^3[3(\ddl^2+5\ddm^2)+3\dl^4+55\dm^4
+6(10\dm^2\ddm+\dl^2\ddl)+3\dl\dm (\dl\dm+2\dl^2+2\ddl )]\, ,\nonumber\\
\eea
and
\bea
\label{Fdef}
F&=&-\ddl +\dl\dps +\frac 14 \zeta 
(3)\{\dl^2[6\dl^4+3\dl^2\dm^2+11\dl^2\ddl+6\dl(\dm^3+\dm\ddm)+
3(2\ddl^2+\ddm^2+2\dm^4 \nonumber\\
& &+3\dm^2\ddm+\dm^2\ddl)]+\ddl(\ddl^2+3\ddm^2+3\dm^4+6\dm^2\ddm)+6\dm^3\dl(\ddl+\ddm)
+3\dl\dm^5+3\ddm\dm\dl(\ddm+2\ddl)\}  \nonumber\\
& & \times \{\dps^2+3\dl^2+6\dm^2\}+
\frac 16(-3\dl^2-6\dm^2+\dps^2+\ap{}^3  h_1)- \ap{}^3  h_5\,, \\
\label{Gdef}
G&=&-\ddm+\dm\dps+\frac 18 \zeta (3)
\{\dm^2[30\dm^4+3\dl^2\dm^2+55\dm^2\ddm+6\dm(\dl^3+\dl\ddl)+
3(10\ddm^2+\ddl^2+2\dl^4 \nonumber\\
& &+3\dl^2\ddl+\dl^2\ddm)]
+\ddm(5\ddm^2+3\ddl^2+3\dl^4+6\dl^2\ddl)+6\dl^3\dm(\ddl+\ddm)
+3\dm\dl^5+3\ddl\dl\dm(\ddl+2\ddm)\} \nonumber\\
& & \times \{\dps^2+3\dl^2+6\dm^2\}-\ap{}^3  h_6\,.
\eea
Here $h_{5}$ and $h_{6}$ are defined in the Appendix.

%%%%%%%%%%%%%%%%%%%%%%%%%%%%%%%%%%%%
\begin{figure}[t]
\begin{center}
\epsfxsize = 3.5in \epsffile{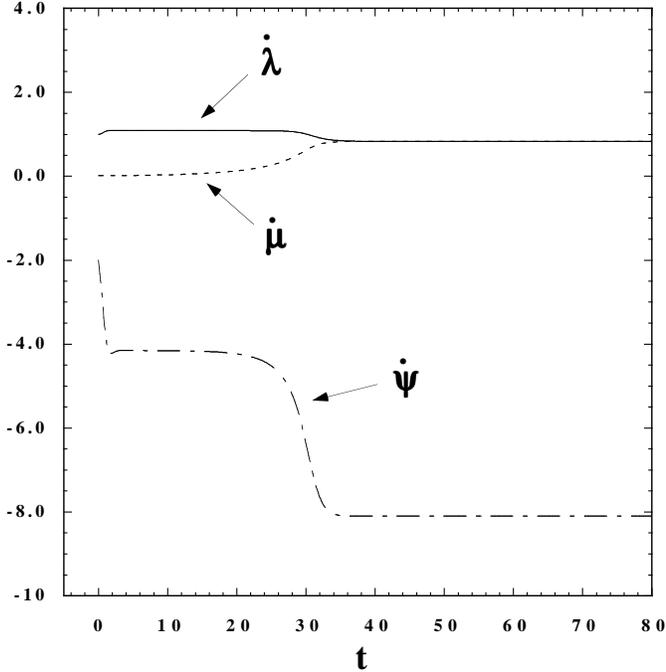} 
\caption{\label{dlanddmu} \small Evolution of $\dot{\lambda}$, $\dot{\mu}$
and $\dot{\psi}$ for the initial condition 
$\dot{\lambda}=1$, $\ddot{\lambda}=0$, 
$\lambda^{(3)}=0$, 
$\dot{\mu}=0.01$, $\ddot{\mu}=0$, $\mu^{(3)}=0$, and 
$\dot{\psi}=-2$.
The solution first approaches the fixed point 
$(\dot{\lambda}, \dot{\mu}, \dot{\psi})=
(1.0852, 0, -4.1566)$ temporaly, but the final attractor 
is the de Sitter fixed point 
$(\dot{\lambda}, \dot{\mu}, \dot{\psi})=
(0.8282, 0.8282, -8.1043)$. }
\end{center}
\end{figure}
%%%%%%%%%%%%%%%%%%%%%%%%%%%%%%%%%

We numerically solve Eqs.~(\ref{ddotpsi}) and (\ref{lammu4})
 to understand the attractor behavior of the system.
In Figure \ref{dlanddmu}  we plot one example for the evolution of 
$\dot{\lambda}$, $\dot{\mu}$ and $\dot{\psi}$.
In this case the initial conditions are  chosen as  
$(\dot{\lambda}_{i}, \dot{\mu}_{i}, \dot{\psi}_{i})=(1, 0.01, -2)$.
We find that the solution temporarily approaches the third fixed point 
in Eq.~(\ref{ideal}) in which only the large dimensions grow exponentially.
However the solution finally approaches the first fixed point 
in Eq.~(\ref{fixedls}) in which both large and small dimensions
exhibit accelerated expansion.
This suggests that the ideal critical point in which large dimensions 
grow but small dimensions are stabilized is not a stable attractor.
In fact we have chosen initial conditions close to the third fixed point 
in Eq.~(\ref{ideal}) and found that solutions repel away from it.
This property also applies for other three fixed points
in Eq.~(\ref{ideal}).

Before closing this section, let us comment on the consistency of the perturbative solutions that we found using the  ansatz Eq.~(\ref{ansatz}) and the fixed points analysis above. We found that there exist de Sitter fixed points that do not exist when higher curvature terms are neglected. These attractor solutions require curvature to be large and correspond to non perturbative solutions in $\alpha^\prime$. On the other hand, at late times our perturbative solutions interpolate to the solutions without corrections and ultimately will follow the attractor $\dl=-w\dps$. The  reason for that is that our solutions will feel the effect of curvature at initial times, then as the universe expands and curvature becomes smaller, they relax to the unperturbed solutions.

\section{Conclusions}

We have studied the effect of $\alpha^\prime$ corrections on string gas cosmology. Starting from
the low energy effective action of type IIB strings including leading order $O( \alpha^\prime {}^3)$ corrections, 
in the $\sigma$-parametrization, we derived the modified equations of motion and studied the adiabatic evolution of the universe. With the assumptions of an initially homogeneous toroidal universe and in thermal equilibrium, filled with an ideal gas of closed strings at weak coupling, we studied the effect of higher curvature corrections finding perturbative (in $\ap{}^3$) solutions and compared them to the uncorrected solutions previously studied in \eg \cite{BBST}.

We considered two cases: the isotropic case, where all dimensions are taken to be equal and the
anisotropic where we consider 3 large and 6 small dimensions.
For the isotropic case we considered two regimes: a) a high energy (temperature) regime, Hagedorn regime,
where the gas of strings is described by a pressureless dust equation of state; b) a low temperature regime,  radiation regime, where the gas of strings is described by a radiation-like equation of state. 
In the Hagedorn case, we found that the correction term acts as a driving force, such that the expansion rate increases compared to the uncorrected case. On the other hand, in the  radiation regime we found that the correction term has the opposite effect, it acts as an initial damping force such that the expansion rate decreases faster. We found that in general the effect of the corrections is small and it is only noticeable at early times. Moreover  it was shown that this effect is stronger in the Hagedorn case than in the   radiation one. We also showed the existence of fixed points, with and without dependence on the equation of state and studied their stability.  We found a stable de Sitter fixed point, which corresponds to a non perturbative solution, for the radiation-like equation of state that leads to geometrical inflation. However one expects that at such large value of $\dot\lambda$ next to leading higher curvature terms must be taken into account as well.

In the anisotropic case, in the  radiation regime, we found that the correction terms act as an initial damping force for both large and small dimensions initially decreasing their expansion rate faster compared to the uncorrected case. Even though the effect is really small, we found that the effect is weaker for the small dimensions. This tells us that at least to order $O(\alpha^\prime{}^3)$  the inclusion of higher curvature terms do not alter the stabilization of the small dimensions of the unperturbed solution for small values of $\dl$. Although the analysis was carried assuming 3 large and 6 small dimensions, 
we expect these features to survive in the more general case of $d$ large and $9-d$ dimensions. We also  
analysed the fixed points of the exact system finding one in which the large dimensions expand whereas the small dimensions are kept constant but is not a stable attractor. On the other hand,  we found a stable de Sitter fixed point which leads to an accelerated expansion of both large and small dimensions.
Once again, as in the isotropic case, this fixed point corresponds to a non perturbative solution.

Our results show that the inclusion of leading $\alpha'$ corrections affects the dynamics only for reasonable high curvature initial configurations. In such a situation, the next to leading terms are expected to become relevant as well. On the other hand, our perturbative approach is only valid at regimes where supergravity description is valid \ie small curvature.  Therefore our work has to be understood as a modest qualitative study of the effect of higher curvature in the BV scenario, to see in which way the dynamics is affected. There are some open issues that deserve further study. For instance,   it is  interesting to perform a detailed study of the prediction of the dimensionality of space-time using our framework \cite{BB}.  

%%%%%%%%%%%%%%%%%%%%%%%%%%%%%%%%%%%%%%%%%%%%%%%
\vskip 25pt
\noindent
{\Large \bf Acknowledgements}
\vskip 10pt

We  thank Shinji Tsujikawa for his collaboration in early stages of the work.
We also thank Matthias Blau, Martin O'Loughlin and Marco Serone for useful discussions, comments and careful
reading of the draft.  We are also grateful to Paolo Creminelli and Jean Pierre Derendinger 
 for useful discussions.
The research of M.\,B. has been supported by the Swiss National Science Foundation and
by the Commission of the European Communities under contract MRTN-CT-2004-005104.

%%%%%%%%%%%%%%%%%%%%%%%%%%%%%%%%%%%%%%%%%%%%%%%
\newpage

\appendix
\section*{Appendix: Explicit forms of $\alpha^{\prime}$ corrections 
in the anisotropic case}
\renewcommand{\theequation}{A-\arabic{equation}}
\setcounter{equation}{0} 

In Eqs.~(\ref{bes1}-\ref{bes4}) the $\alpha^{\prime}$  are given by 

\begin{eqnarray}
\label{h1corre}
h_{1}&=&\frac18 \zeta (3)
  \Big[504  {{{\dl} }^8}+  1260{{{\dl} }^6}
{{{\dm} }^2}+1512 
{{{\dl} }^5}  {{{\dm} }^3}+   
   1638  {{{\dl} }^4}  {{{\dm} }^4}+3024
{{{\dl} }^3}  {{{\dm} }^5}+2772 
{{{\dl} }^2}  {{{\dm} }^6}+   4410  {{{\dm} }^8} 
+144  {{{\dl} }^7} \dps  \nonumber \\ &&
+144  {{{\dl} }^5} {{{\dm} }^2}  {\dps } +  
288  {{{\dl} }^4}  {{{\dm} }^3} \dps  +288  {{{\dl} }^3}
{{{\dm} }^4}  {\dps } +144  {{{\dl} }^2}  {{{\dm} }^5} \dps  + 
720  {{{\dm} }^7}  {\dps }-  216  {{{\dl} }^4}  {{{\dm} }^2}
{\ddl} -144  {{{\dl} }^3} {{{\dm} }^3}  {\ddl} + 144  {{{\dl} }^2}  {{{\dm} }^4}
{\ddl} \nonumber \\ &&
+216  {\dl} {{{\dm} }^5}  {\ddl} + 264  {{{\dl} }^5} {\dps }{\ddl} + 216 {{{\dl} }^3}  {{{\dm} }^2} \dps  {\ddl} + 360  {{{\dl} }^2}  {{{\dm} }^3} \dps    {\ddl} +216 {\dl}{{{\dm} }^4}  \dps \ddl-  396  {{{\dl} }^4}  {{{\ddl}
}^2}-396  {{{\dl} }^2}  {{{\dm} }^2} 
 {{{\ddl} }^2}- 216  {\dl}   {{{\dm} }^3}  {{\ddl }^2}\nonumber \\ &&
+36  {{{\dm} }^4} {{{\ddl} }^2}+  144  {{{\dl} }^3}  {\dps }
{{{\ddl} }^2}+ 72 {\dl}   {{{\dm} }^2}  {\dps }
{{{\ddl} }^2}+  72  {{{\dm} }^3}  {\dps }   {{\ddl }^2}-96 
{{{\dl} }^2}  {{{\ddl} }^3}-72
{{{\dm} }^2}  {{{\ddl} }^3}+
   24  {\dl}   {\dps }   {{\ddl }^3}+18  {{{\ddl} }^4}+216  {{{\dl} }^5}  {\dm}
\ddm  \nonumber \\ &&
+144  {{{\dl} }^4}  {{{\dm} }^2} \ddm-144 {{{\dl} }^3}  {{{\dm} }^3}  \ddm -  
   216  {{{\dl} }^2}  {{{\dm} }^4} \ddm+216  {{{\dl} }^4} 
 {\dm}   {\dps }   \ddm  +  360  {{{\dl} }^3}  {{{\dm} }^2}
    \dps    \ddm +216  {{{\dl} }^2}  {{{\dm} }^3}
{\dps }   \ddm +  
   1320  {{{\dm} }^5}  {\dps }  \ddm - 288 {{{\dl} }^3}  {\dm}   {\ddl}   \ddm \nonumber \\ &&  
 -  648  {{{\dl} }^2}  {{{\dm} }^2}
{\ddl}   \ddm- 288  {\dl}   {{{\dm} }^3}
{\ddl}   \ddm+   
   288  {{{\dl} }^2}  {\dm}  \dps    \ddl  \ddm + 
   288  {\dl}   {{{\dm} }^2} \dps    \ddl  \ddm - 
   216  {\dl}   {\dm}   {{{\ddl} }^2}  \ddm+72  {\dm}  \dps    
   {{\ddl }^2}  \ddm +  
   36  {{{\dl} }^4}  {{\ddm}^2}- 216  {{{\dl} }^3}  {\dm}  
    {{\ddm }^2} \nonumber \\ && 
   -396  {{{\dl} }^2}  {{{\dm} }^2}  {{\ddm }^2}-1980  {{{\dm} }^4}  {{\ddm }^2}
+      72  {{{\dl} }^3}  {\dps }   {{\ddm }^2}+ 
   72 {{{\dl} }^2}  {\dm}   {\dps }
{{\ddm }^2}+   
   720  {{{\dm} }^3}  {\dps }   {{\ddm }^2}-216 
{\dl}   {\dm}   {\ddl}
{{\ddm }^2}+ 
   72  {\dl}   {\dps }   \ddl   {{\ddm}^2} + 108  {{{\ddl} }^2}  {{\ddm }^2}\nonumber \\ &&   
 -  72  {{{\dl} }^2}  {{\ddm}^3}
-480  {{{\dm} }^2}  {{\ddm}^3}+ 
120  {\dm}   {\dps }
{{\ddm }^3}+  
   90  {{\ddm }^4}-264  {{{\dl}}^5}  {\lambda^{(3)}} - 
216  {{{\dl} }^3} {{{\dm} }^2}  {\lambda^{(3)}} -  360  {{{\dl} }^2}  {{{\dm} }^3} {\lambda^{(3)}} - 216  {\dl}   
{{{\dm} }^4}  {\lambda^{(3)}}  \nonumber \\ &&  
  - 288  {{{\dl} }^3}  {\ddl}
{\lambda^{(3)}} -  
144  {\dl}  
{{{\dm} }^2}  {\ddl}
{\lambda^{(3)}} - 
   144  {{{\dm} }^3}  {\ddl}
{\lambda^{(3)}} -72  {\dl}  
{{{\ddl} }^2}  {\lambda^{(3)}} -
 288  {{{\dl} }^2}  {\dm}   \ddm  {\lambda^{(3)}} - 288 
{\dl}   {{{\dm} }^2}  \ddm  {\lambda^{(3)}} - 144  {\dm}   {\ddl}  \ddm  {\lambda^{(3)}} \nonumber \\ &&
  - 72 {\dl}   {{\ddm }^2}
{\lambda^{(3)}} -  216  {{{\dl} }^4}  {\dm}   {\mu^{(3)}}
-360  {{{\dl} }^3}  {{{\dm} }^2}  {\mu^{(3)}} -   
   216  {{{\dl} }^2}  {{{\dm} }^3}
{\mu^{(3)}} -1320  {{{\dm} }^5}  {\mu^{(3)}} - 
  288  {{{\dl} }^2}  {\dm}   \ddl   {\mu^{(3)}}  - 288 {\dl}   {{{\dm} }^2}  {\ddl}   {\mu^{(3)}} \nonumber \\ &&  
  - 72  {\dm}   {{{\ddl} }^2}
{\mu^{(3)}} -144  {{{\dl} }^3} \ddm   {\mu^{(3)}} -  144  {{{\dl} }^2}  {\dm}   \ddm  {\mu^{(3)}} -1440 
{{{\dm} }^3}  \ddm   {\mu^{(3)}} -  144  {\dl}   {\ddl}  \ddm  {\mu^{(3)}} - 360 {\dm}   {{\ddm }^2}  {\mu^{(3)}}\Big], 
\end{eqnarray}
\begin{eqnarray}
h_{2}&=&\frac 12 \zeta (3)  \Big[24  {{{\dl} }^7} 
\dps  -33  {{{\dm} }^6}  {\ddl} -3  {{{\dm} }^5}  \dps 
{\ddl} +  {{\dps  }^2}  {{{\ddl}}^3}+8  {{\ddl}^4}+3  {{\dps  }^2}  \ddl  {{\ddm }^2}+
  9  {{{\ddl} }^2}  {{\ddm}^2}+15  {\ddl}   {{\ddm }^3}+ 9  {{{\dl} }^5}  
  \big(5  {{{\dm} }^2}  \dps  -4 \dps   
{\ddl} \nonumber \\ &&
-10  {\dm}   \ddm\big)+  6  {{{\dl} }^6}  \big({{\dps 
}^2}-28  {\ddl} -\ddps   \big)
-{{{\ddl} }^3}  \ddps -  3  {\ddl}   {{\ddm }^2}
\ddps   -6  \dps    {{\ddl}^2}  {\lambda^{(3)}} -  
  6  \dps    {{\ddm }^2}
{\lambda^{(3)}} +  6  {\ddl}   {{{\lambda^{(3)}} }^2} - 12  \dps    {\ddl}  \ddm 
{\mu^{(3)}} \nonumber \\ &&+12  \ddm   {\lambda^{(3)}}
{\mu^{(3)}} + 6  {\ddl}   {{{\mu^{(3)}} }^2}+
 3  {\dm}   \big[-3  \dps    \ddl  
\ddm   ({\ddl} +3 \ddm )+   8  {{\ddm }^2}  {\lambda^{(3)}} +  3
{{{\ddl} }^2}  {\mu^{(3)}}  2  {\ddl}   \ddm
\big(2  {\lambda^{(3)}} +7 
{\mu^{(3)}} \big)\big]\nonumber \\ &&-  
  3  {{{\dm} }^3}    
  \big[\dps    {\ddl}   (3
{\ddl} +10 
\ddm )-  
  2  \big(4  \ddm   {\lambda^{(3)}} +\ddl   \big(2 
{\lambda^{(3)}} +3  {\mu^{(3)}} \big)\big)\big]+  3  {{{\ddl} }^2}  {\lambda^{(4)}} +3
{{\ddm }^2}  {\lambda^{(4)}} + 3{{{\dm} }^4}  \big[{{\dps  }^2}
{\ddl} +{\ddl}   (3  \ddm -\ddps )
\nonumber \\ &&
- 2 \dps {\lambda^{(3)}} +{\lambda^{(4)}} \big]+ {{{\dl} }^4}  \big[45  {{{\dm} }^3} \dps +11  {{\dps  }^2}  {\ddl} -9 {{\ddm}^2}+ 3 {{{\dm} }^2}  \big({{\dps  }^2}-75
{\ddl} -45 \ddm -\ddps   \big)-11  {\ddl}   {\ddps }
 -22  {\dps }   {\lambda^{(3)}} \nonumber \\ &&+ 
  3  {\dm}   \big(\dps    \ddm-3  {\mu^{(3)}} \big)+11 
{\lambda^{(4)}} \big]+   6  {\ddl}   \ddm
{\mu^{(4)}} + {{{\dl} }^3}  \big[39  {{{\dm} }^4} 
  \dps -6 \dps \big(11  {{{\ddl} }^2}+{{\ddm}^2}\big)+  6  {{{\dm} }^3}  \big({{\dps  }^2}-30
{\ddl} -26 \ddm -\ddps   \big)\nonumber \\ &&
+  88  {\ddl}   {\lambda^{(3)}} +6 \ddm  {\mu^{(3)}} -6  {{{\dm} }^2}  \big(2  \dps 
({\ddl} +\ddm)+  {\mu^{(3)}} \big)+ 6  {\dm}   \big({{\dps  }^2} \ddm-6 
{\ddl}   \ddm -2{{\ddm }^2}- \ddm  \ddps   -2 \dps    {\mu^{(3)}} +{\mu^{(4)}} \big)\big]\nonumber \\ &&
+  3  {{{\dm} }^2}  \big[{{{\ddl} }^3}+ 9  {{{\ddl} }^2}  \ddm-4  \dps    \ddm
{\lambda^{(3)}} +   4  {\lambda^{(3)}}   {\mu^{(3)}} +2  \ddm
{\lambda^{(4)}} + {\ddl}   \big(2  {{\dps}^2}  \ddm+17  {{\ddm }^2}-2  \ddm \ddps   - 
  4  \dps    {\mu^{(3)}} +2  {\mu^{(4)}} \big)\big]\nonumber \\ &&
+ 3  {{{\dl} }^2} \big\{18  {{{\dm} }^5}  \dps  +22 {{{\ddl} }^3}+2 {\ddl}   {{\ddm }^2}+ 3 {{\ddm }^3}+{{\dps  }^2}
\big(2  {{{\ddl} }^2}+{{\ddm }^2}\big)- 2  {{{\ddl} }^2} \ddps   -{{\ddm}^2} \ddps   + 
  {{{\dm} }^4}  \big(2  {{\dps  }^2}-39 {\ddl} - 90  \ddm +\ddps  \big)  
\nonumber \\ &&
+4  {{{\lambda^{(3)}} }^2}-    2  {{{\dm} }^3}  \big(\dps    (3
{\ddl} +5 \ddm )-{\mu^{(3)}} \big)+  2 {{{\mu^{(3)}} }^2}-4  \dps    \big(2 {\ddl}  
{\lambda^{(3)}} + \ddm   {\mu^{(3)}} \big)+  {\dm}   \big[-\dps    \ddm  (10  {\ddl} +9  \ddm )+  
  4  \big(2  {\ddl}   {\mu^{(3)}} \nonumber \\ &&
+\ddm   \big({\lambda^{(3)}} +3 {\mu^{(3)}} \big)\big)\big]+   4  {\ddl}   {\lambda^{(4)}} +2 \ddm  {\mu^{(4)}} +
  {{{\dm} }^2}  \big[6  {{\ddm }^2}+ {{\dps  }^2} ({\ddl} +3  \ddm )-   3  \ddm  \ddps  
-\ddl ( 6  \ddm +\ddps )-   2  \dps    \big({\lambda^{(3)}} +3  {\mu^{(3)}}
\big)\nonumber \\ &&
+{\lambda^{(4)}} +3  {\mu^{(4)}} \big]\big\}+  {\dl}     
  \big\{33  {{{\dm} }^6}  \dps - 2  \dps    \big(10  {{\ddl}^3}+9  {\ddl}   {{\ddm }^2}+3  {{\ddm}^3}\big)+  
  3  {{{\dm} }^5}  \big({{\dps  }^2}-36
{\ddl} -66 \ddm -\ddps   \big)-3  {{{\dm} }^4}  \big(4  \dps ({\ddl} +
2 \ddm )\nonumber \\ && -3  {\mu^{(3)}} \big)+   3  \big(16  {{{\ddl} }^2}  {\lambda^{(3)}}
+10  {\ddl} \ddm   {\mu^{(3)}} +  {{\ddm }^2}  \big(4  {\lambda^{(3)}} +5
{\mu^{(3)}} \big)\big)+   {{{\dm} }^2}  \big[-3  \dps    \big(3  {{{\ddl} }^2}+18  {\ddl}   \ddm +11  {{\ddm}^2}\big)
+ 12  \big({\ddl}   \big({\lambda^{(3)}} \nonumber \\ &&
+3 {\mu^{(3)}} \big)+  
  \ddm   \big(3  {\lambda^{(3)}} +4  {\mu^{(3)}} \big)\big)\big]+  
  6  {{{\dm} }^3}  \big[6  {{\ddm}^2}+{{\dps  }^2}  ({\ddl} +\ddm )+  
  {\ddl}   (2  \ddm -\ddps  )-\ddm  \ddps  -2  \dps    \big({\lambda^{(3)}} +{\mu^{(3)}}
\big)+{\lambda^{(4)}} +{\mu^{(4)}} \big]\nonumber \\ &&+ 3  {\dm}   \big[6  {{{\ddl} }^2}
\ddm +10 {{\ddm }^3}+  {{\dps  }^2}  \ddm   (2
\ddl+\ddm )-{{\ddm }^2} \ddps +  4  {\lambda^{(3)}}   {\mu^{(3)}} + 
  2  {{{\mu^{(3)}} }^2}- 4  \dps \big({\ddl}   {\mu^{(3)}} +\ddm  \big({\lambda^{(3)}} +{\mu^{(3)}} \big)\big)+
  2  \ddm   {\lambda^{(4)}} \nonumber \\ &&
+2  \ddm  {\mu^{(4)}} +
  2  {\ddl}   \big(9  {{\ddm}^2}-\ddm \ddps   +{\mu^{(4)}}
\big)\big]\big\}\Big],
\end{eqnarray}
\begin{eqnarray}
h_3&=&\frac 14 \zeta (3) \Big[210 {{{\dm}}^7}\dps +15 {{{\dl}}^6} ({\dm}
\dps -\ddm )-  
 180 {{{\dm}}^5}\dps  \ddm
+3 {{\dps }^2} {{\ddl}^2} \ddm +
15 {{\ddl}^3} \ddm +9
{{\ddl}^2} {{\ddm }^2}+5 {{\dps }^2}
{{\ddm }^3}+  
 40 {{\ddm }^4} \nonumber\\ &&+3 {{{\dl}}^5} (9
{{{\dm}}^2}
\dps -\dps  \ddm +
 {\dm} ({{\dps }^2}-30 \ddl-18 \ddm -\ddps  ))+  
 30 {{{\dm}}^6} ({{\dps }^2}-49
\ddm -\ddps  )-3 {{\ddl}^2} \ddm 
\ddps  - 
 5 {{\ddm }^3} \ddps  -12
\dps  \ddl \ddm 
{\lambda^{(3)}}\nonumber\\ &&+  
6 \ddm {{{\lambda^{(3)}}}^2}-6 \dps 
{{\ddl}^2} {\mu^{(3)}}-  
 30\dps  {{\ddm }^2}
{\mu^{(3)}}+12 \ddl {\lambda^{(3)}} {\mu^{(3)}}+ 
 30 \ddm {{{\mu^{(3)}}}^2}+ 
 {{{\dm}}^3}
(-6 \dps ({{\ddl}^2}+55 {{\ddm}^2})+  
 6 \ddl {\lambda^{(3)}} \nonumber\\ && +440 \ddm {\mu^{(3)}})+
 {\dm} (-2\dps  (3 {{\ddl}^3}+9
{{\ddl}^2} \ddm +50
{{\ddm }^3})+ 
 3 (10 \ddl \ddm 
{\lambda^{(3)}}+80 {{\ddm }^2} {\mu^{(3)}}+  
 {{\ddl}^2} (5 {\lambda^{(3)}}+4
{\mu^{(3)}})))\nonumber\\ &&+ 
 6 \ddl \ddm 
{\lambda^{(4)}}+3 {{\ddl}^2} {\mu^{(4)}}+15 {{\ddm }^2}
{\mu^{(4)}}+  
 3 {{{\dl}}^4} [13 {{{\dm}}^3}
\dps +{\dps ^2} \ddm +3 \ddl
\ddm + 
 {{{\dm}}^2} (2 {{\dps }^2}-45
\ddl-39
\ddm -2 \ddps  )\nonumber\\ &&-\ddm 
 \ddps  +{\dm} (-4 \dps 
(2 \ddl+\ddm )+  
3 {\lambda^{(3)}})- 
 2\dps  {\mu^{(3)}}+{\mu^{(4)}}]+  
 {{{\dm}}^4} [-9 {{\ddl}^2}+55
({{\dps }^2}
\ddm -\ddm \ddps  -  
 2\dps  {\mu^{(3)}}+{\mu^{(4)}})]  \nonumber\\ &&
 +3 {{{\dl}}^3} [30 {{{\dm}}^4}
\dps -\dps \ddm (10 \ddl+3
\ddm )+
 2 {{{\dm}}^3} ({{\dps }^2}-26
\ddl-60
\ddm -\ddps  )-  
 2 {{{\dm}}^2} (\dps  (5 \ddl+3
\ddm )-{\lambda^{(3)}})+ 
 6 \ddm {\lambda^{(3)}}\nonumber\\ &&+8 \ddl {\mu^{(3)}}+4
\ddm {\mu^{(3)}}+  
 2 {\dm} (6 {{\ddl}^2}+{{\dps }^2}
(\ddl+\ddm )+
 \ddl (2 \ddm
-\ddps  )-\ddm \ddps  -  
 2\dps 
({\lambda^{(3)}}+
{\mu^{(3)}})+{\lambda^{(4)}}+{\mu^{(4)}})]\nonumber\\ &&+
 3 {{{\dl}}^2} \{33 {{{\dm}}^5}
\dps +2 {\dps ^2} \ddl \ddm +  
 17 {{\ddl}^2} \ddm +9
\ddl {{\ddm }^2}+{{\ddm }^3}+
 {{{\dm}}^4} ({{\dps }^2}-
 90\ddl-
 165 \ddm -\ddps  )-   2 \ddl \ddm 
\ddps  -   2 {{{\dm}}^3} (2\dps  (\ddl+\ddm )\nonumber\\ &&+{\lambda^{(3)}})+  
 4 {\lambda^{(3)}} {\mu^{(3)}}-4\dps  (\ddm 
{\lambda^{(3)}}+\ddl {\mu^{(3)}})+
 {\dm} [-\dps  (11 {{\ddl}^2}+ 
 18\ddl \ddm +3
{{\ddm }^2})+  4 (\ddm (3
{\lambda^{(3)}}+{\mu^{(3)}})+\ddl
  (4 {\lambda^{(3)}}\nonumber\\ &&+3 {\mu^{(3)}}))]+2 \ddm 
{\lambda^{(4)}}+2 \ddl  {\mu^{(4)}}+
 {{{\dm}}^2} [6 {{\ddl}^2}+
 {{\dps }^2}
(3 \ddl+\ddm )-  
 \ddm \ddps  -3
\ddl (2 \ddm +\ddps  )- 
 2\dps  (3 {\lambda^{(3)}}+{\mu^{(3)}})\nonumber\\ &&+3
{\lambda^{(4)}}+{\mu^{(4)}}]\} +
 3 {{{\dm}}^2} \big\{3 {{\ddl}^3}+{{\dps }^2}
({{\ddl}^2}+10 {{\ddm}^2})+   
 {{\ddl}^2} (2 \ddm -\ddps  )-  
 4\dps  (\ddl
{\lambda^{(3)}}+10
 \ddm  {\mu^{(3)}})+  
 2 \ddl {\lambda^{(4)}}\nonumber\\ &&+2 \big(55
{{\ddm }^3}-5
{{\ddm }^2} \ddps  +
 {{{\lambda^{(3)}}}^2}+10 {{{\mu^{(3)}}}^2}+10 \ddm {\mu^{(4)}}\big)\big\}- 
 3 {\dl} \big\{66 {{{\dm}}^5}
\ddl+9 \dps {{\ddl}^2} \ddm +
3\dps  \ddl {{\ddm }^2}-14
\ddl \ddm 
{\lambda^{(3)}}  
\nonumber\\ && -3 {{\ddm }^2} {\lambda^{(3)}}+{{{\dm}}^4}
  (-\dps \ddl+3 {\lambda^{(3)}})-  8 {{\ddl}^2} {\mu^{(3)}}-4 \ddl \ddm  {\mu^{(3)}}+   
 {{{\dm}}^2} [\dps  \ddl (9 \ddl+10 \ddm )-  
  4 (2 \ddm {\lambda^{(3)}}+\ddl (3 {\lambda^{(3)}}+{\mu^{(3)}}))]\nonumber\\ &&-
 2 {{{\dm}}^3} [{{\dps }^2}
\ddl-2
{{\ddl}^2}- 
 \ddl (6 \ddm
+\ddps  )-2\dps  {\lambda^{(3)}}+{\lambda^{(4)}}]-  
 {\dm} \big[10 {{\ddl}^3}+{{\dps }^2} 
 \ddl (\ddl+2 \ddm )+
 {{\ddl}^2} (18 \ddm-
\ddps  )\nonumber\\ &&-  
 4\dps  (\ddm 
{\lambda^{(3)}}+\ddl ({\lambda^{(3)}}+{\mu^{(3)}}))+  
 2 \big({{{\lambda^{(3)}}}^2}+2 {\lambda^{(3)}}
{\mu^{(3)}}+
\ddm {\lambda^{(4)}}\big)+ 2 \ddl (3 {{\ddm}^2}
-\ddm
 \ddps  +{\lambda^{(4)}}+{\mu^{(4)}})\big]\big\}\Big],
\end{eqnarray}
\begin{eqnarray}
\label{h4corre}
 h_4&=&\frac{3}{2}  \zeta (3) \Big[24  \dl ^8+210  \dm ^8+6  \dl ^7 \dps +30  \dm ^7 \dps + 
  \ddl ^4+6  \dl ^6 (10  \dm ^2+ \ddl )+30  \dm ^6  \ddm   + 
     6  \ddl ^2  \ddm ^2+5  \ddm ^4\nonumber\\&&+
      \dm ^4 (3  \ddl ^2-55  \ddm ^2)+
  \dm ^2 (-3  \ddl ^3+3  \ddl ^2  \ddm -10  \ddm ^3)+  
   \dl ^5 (72  \dm ^3+6  \dm ^2 \dps +11 \dps   \ddl + 
 12  \dm   \ddm -11 \mu ^{(3)} )  \nonumber\\&&+
  \dl ^4 (78  \dm ^4+12  \dm ^3 \dps -11  \ddl ^2- 
 6  \dm ^2 ( \ddl -2  \ddm )+3  \ddm ^2 + 
  9  \dm  (\dps   \ddm -\nu ^{(3)} ))+ 
 55  \dm ^5 (\dps   \ddm -\nu ^{(3)} )  \nonumber\\&&+
  3  \dl ^3 (48  \dm ^5+4  \dm ^4 \dps -2  \dm   \ddm  ( \ddl + \ddm )+ 
 \dps  (2  \ddl ^2+ \ddm ^2) + 
   \dm ^2 (\dps  (3  \ddl +5  \ddm )-3 \mu ^{(3)} -
 5 \nu ^{(3)} )\nonumber\\&&-
 2 (2  \ddl  \mu ^{(3)} + \ddm  \nu ^{(3)} )) + 
  3  \dm ^3  
 (\dps  ( \ddl ^2+10  \ddm ^2)-
 2 ( \ddl  \mu ^{(3)} +10  \ddm  \nu ^{(3)} ))+ 
   \dm  (\dps  (3  \ddl ^2  \ddm +5  \ddm ^3)  \nonumber\\&&-
 3 (2  \ddl   \ddm  \mu ^{(3)} + \ddl ^2 \nu ^{(3)}  + 
  5  \ddm ^2 \nu ^{(3)} ))+
  \dl ^2 (132  \dm ^6+6  \dm ^5 \dps -2  \ddl ^3 + 
  6  \dm ^4 (2  \ddl - \ddm )+3  \ddl   \ddm ^2-
 3  \ddm ^3 \nonumber\\&&-
 3  \dm ^2 (5  \ddl ^2+6  \ddl   \ddm +5  \ddm ^2)+ 
 3  \dm ^3 (\dps  (5  \ddl +3  \ddm )-5 \mu ^{(3)}  -
  3 \nu ^{(3)} )+
 3  \dm  (\dps   \ddm  (4  \ddl + \ddm ) -
  2 (2  \ddl  \nu ^{(3)} \nonumber\\ &&+ \ddm  (2 \mu ^{(3)} +\nu ^{(3)} ))))+
  \dl  (12  \dm ^5  \ddl -6  \dm ^3  \ddl  ( \ddl + \ddm ) -
  6  \dm   \ddl   \ddm  ( \ddl + \ddm )+
 \dps  ( \ddl ^3+3  \ddl   \ddm ^2)  -
  9  \dm ^4 (\dps   \ddl \nonumber\\ &&-\mu ^{(3)} )- 
 3 ( \ddl ^2 \mu ^{(3)} + \ddm ^2 \mu ^{(3)}   +
  2  \ddl   \ddm  \nu ^{(3)} )+
 3  \dm ^2 (\dps   \ddl  ( \ddl +4  \ddm ) -
  2 (2  \ddm  \mu ^{(3)} + \ddl  (\mu ^{(3)} +2 \nu ^{(3)} ))))\Big]\nonumber\\
\end{eqnarray}
We checked that these expressions reduce to those in 
the isotropic case by setting $\lambda=\mu$.

In Eqs.~(\ref{Fdef}) and (\ref{Gdef}) the $h_{5}$ and 
$h_{6}$ terms are defined by 
\bea
h_5&=& \zeta (3) \frac 12 \Big[24  {{{\dl} }^7} 
\dps  -33  {{{\dm} }^6}  {\ddl} -3  {{{\dm} }^5}  \dps 
{\ddl} + 
  {{\dps  }^2}  {{{\ddl}}^3}+8  {{\ddl}^4}+3  {{\dps  }^2}  \ddl  {{\ddm }^2}+
  9  {{{\ddl} }^2}  {{\ddm}^2}+15  {\ddl}   {{\ddm }^3}+
  9  {{{\dl} }^5}  
  \big(5  {{{\dm} }^2}  \dps  -4 \dps   
{\ddl} \nonumber\\ &&-10  {\dm}   \ddm\big)+  
  6  {{{\dl} }^6}  \big({{\dps 
}^2}-28  {\ddl}    \big)
 -    6  \dps    {{\ddl}^2}  {\lambda^{(3)}} -  
  6  \dps    {{\ddm }^2}
{\lambda^{(3)}} +6  {\ddl}   {{{\lambda^{(3)}} }^2}- 
  12  \dps    {\ddl}  \ddm 
{\mu^{(3)}} +12  \ddm   {\lambda^{(3)}}
{\mu^{(3)}} +  
  6  {\ddl}   {{{\mu^{(3)}} }^2}\nonumber\\ &&+
  3  {\dm}   \big[-3  \dps    \ddl  
\ddm   ({\ddl} +3 \ddm )+  
  8  {{\ddm }^2}  {\lambda^{(3)}} +3
{{{\ddl} }^2}  {\mu^{(3)}} 
  2  {\ddl}   \ddm
\big(2  {\lambda^{(3)}} +7 
{\mu^{(3)}} \big)\big]-  
  3  {{{\dm} }^3}    
  \big[\dps    {\ddl}   (3{\ddl} +10 
\ddm )-  
  2  \big(4  \ddm   {\lambda^{(3)}} \nonumber\\ &&+\ddl   \big(2 
{\lambda^{(3)}} +3  {\mu^{(3)}} \big)\big)\big]+  
     3{{{\dm} }^4}  \big[{{\dps  }^2}
{\ddl} +3{\ddl}    \ddm   -  
  2  \dps    {\lambda^{(3)}}  \big]+
  {{{\dl} }^4}  \big[45  {{{\dm} }^3} \dps  
+11  {{\dps  }^2}  {\ddl} -9  {{\ddm}^2}+  
  3  {{{\dm} }^2}  \big({{\dps  }^2}-75
{\ddl} \nonumber\\ &&-45 
\ddm  \big)-22  {\dps }   {\lambda^{(3)}} + 
  3  {\dm}   \big(\dps    \ddm-3  {\mu^{(3)}} \big) \big]+ 
  {{{\dl} }^3}  \big[39  {{{\dm} }^4} 
  \dps  -6  \dps    \big(11  {{{\ddl} }^2}+{{\ddm}^2}\big)+ 
  6  {{{\dm} }^3}  \big({{\dps  }^2}-30
{\ddl} -26 
\ddm    \big)  \nonumber \\&&
  +88  {\ddl}   {\lambda^{(3)}} +6 \ddm  {\mu^{(3)}} -
  6  {{{\dm} }^2}  \big(2  \dps 
({\ddl} +\ddm)+{\mu^{(3)}} \big)+  
  6  {\dm}   \big({{\dps  }^2} \ddm-6 
{\ddl}   \ddm -2
{{\ddm }^2}-  
     2 \dps    {\mu^{(3)}}  \big)\big]+  
  3  {{{\dm} }^2}    
  \big[{{{\ddl} }^3}\nonumber \\&& +
  9  {{{\ddl} }^2}  \ddm-4  \dps    \ddm
{\lambda^{(3)}} +  
  4  {\lambda^{(3)}}   {\mu^{(3)}}  +  
  {\ddl}   \big(2  {{\dps 
}^2}  \ddm+17  {{\ddm }^2}   - 
  4  \dps    {\mu^{(3)}}  \big)\big]+
  3  {{{\dl} }^2}    
  \big\{18  {{{\dm} }^5}  \dps  +22
{{{\ddl} }^3}+2 
{\ddl}   {{\ddm }^2} + 3  {{\ddm }^3}\nonumber\\ &&
+{{\dps  }^2} \big(2  {{{\ddl} }^2}+{{\ddm }^2}\big)  + {{{\dm} }^4}  \big(2  {{\dps  }^2}-39
{\ddl} -2 
(45  \ddm  
)\big)+ 
  4  {{{\lambda^{(3)}} }^2}-  
  2  {{{\dm} }^3}  \big(\dps    (3
{\ddl} +5 
\ddm )-{\mu^{(3)}} \big)+ 
  2  {{{\mu^{(3)}} }^2}-4  \dps    \big(2
{\ddl}  
{\lambda^{(3)}}\nonumber\\ && +\ddm   {\mu^{(3)}} \big)+  
  {\dm}   \big[-\dps    \ddm  (10  {\ddl} +9  \ddm )
  4  \big(2  {\ddl}   {\mu^{(3)}} +\ddm   \big({\lambda^{(3)}} +3 
{\mu^{(3)}} \big)\big)\big] +
  {{{\dm} }^2}  \big[6  {{\ddm }^2}+{{\dps  }^2} 
({\ddl} +3  \ddm )
-\ddl  6  \ddm  -  
  2  \dps    \big({\lambda^{(3)}}\nonumber \\&& +3  {\mu^{(3)}}\big) \big]\big\} +  {\dl}     
  \big\{33  {{{\dm} }^6}  \dps  -
  2  \dps    \big(10  {{\ddl}^3}+9  {\ddl}   {{\ddm }^2}+3  {{\ddm}^3}\big)+  
  3  {{{\dm} }^5}  \big({{\dps  }^2}-36
{\ddl} -66 
\ddm \big)-
  3  {{{\dm} }^4}  \big(4  \dps 
({\ddl} +2 
\ddm )-3  {\mu^{(3)}} \big)  \nonumber \\&&+
  3  \big(16  {{{\ddl} }^2}  {\lambda^{(3)}}
+10  {\ddl}  
\ddm   {\mu^{(3)}} +  
  {{\ddm }^2}  \big(4  {\lambda^{(3)}} +5
{\mu^{(3)}} \big)\big)+  
  {{{\dm} }^2}    
  \big[-3  \dps    \big(3  {{{\ddl} }^2}+18  {\ddl}   \ddm +11  {{\ddm}^2}\big)+    
  12  \big({\ddl}   \big({\lambda^{(3)}} + 3
{\mu^{(3)}} \big)\nonumber \\&&+  
  \ddm   \big(3  {\lambda^{(3)}} +4  {\mu^{(3)}}
\big)\big)\big]+
  6  {{{\dm} }^3}  \big[6  {{\ddm}^2}+{{\dps  }^2} 
({\ddl} +\ddm )+  
  2{\ddl}     \ddm     -
  2  \dps    \big({\lambda^{(3)}} +{\mu^{(3)}}
\big)  \big]+ 
  3  {\dm}   \big[6  {{{\ddl} }^2}
\ddm +10 
{{\ddm }^3}+  
  {{\dps  }^2}  \ddm   (2
\ddl\nonumber \\&&+\ddm )+
  4  {\lambda^{(3)}}   {\mu^{(3)}} +2  {{{\mu^{(3)}} }^2}-
  4  \dps    
  \big({\ddl}   {\mu^{(3)}} +\ddm  \big({\lambda^{(3)}} +{\mu^{(3)}} \big)\big)+
  18  {\ddl}     {{\ddm}^2}\big]\big\}\Big]\, ,
\eea
\bea
h_6&=&\frac 14 \zeta (3)\Big[210 {{{\dm}}^7}\dps +15 {{{\dl}}^6} ({\dm}
\dps -\ddm )-  
 180 {{{\dm}}^5}\dps  \ddm
+3 {{\dps }^2} {{\ddl}^2} \ddm +
 15 {{\ddl}^3} \ddm +9
{{\ddl}^2} {{\ddm }^2}+5 {{\dps }^2}
{{\ddm }^3}+  
 40 {{\ddm }^4}+
 3 {{{\dl}}^5} (9
{{{\dm}}^2}
\dps \nonumber \\&&-\dps  \ddm +
 {\dm} ({{\dps }^2}-30 \ddl-18 \ddm ))+  
 30 {{{\dm}}^6} ({{\dps }^2}-49
\ddm   )    -12\dps  \ddl \ddm 
{\lambda^{(3)}}+ 
6 \ddm {{{\lambda^{(3)}}}^2}- 
6 \dps 
{{\ddl}^2} {\mu^{(3)}}- 
 30\dps  {{\ddm }^2}
{\mu^{(3)}}\nonumber \\&&+12 \ddl {\lambda^{(3)}} {\mu^{(3)}}+  
 30 \ddm {{{\mu^{(3)}}}^2}+{{{\dm}}^3}
(-6 \dps ({{\ddl}^2}+55 {{\ddm}^2})+ 
 6 \ddl {\lambda^{(3)}}+440 \ddm {\mu^{(3)}})+
 {\dm} (-2\dps  (3 {{\ddl}^3} +9
{{\ddl}^2} \ddm +50
{{\ddm }^3})\nonumber \\&&+  
 3 (10 \ddl \ddm 
{\lambda^{(3)}}+80 {{\ddm }^2} {\mu^{(3)}}+  
 {{\ddl}^2} (5 {\lambda^{(3)}}+ 
 4{\mu^{(3)}})))+  3 {{{\dl}}^4} [13 {{{\dm}}^3}
\dps +{\dps ^2} \ddm +3 \ddl
\ddm +  {{{\dm}}^2} (2 {{\dps }^2} -45
\ddl-39\ddm  ) \nonumber \\&&+{\dm} (-4 \dps 
(2 \ddl+\ddm )+3
{\lambda^{(3)}})-  
 2\dps  {\mu^{(3)}}]+  
 {{{\dm}}^4} [-9 {{\ddl}^2}+55
({{\dps }^2} \ddm  -   2\dps  {\mu^{(3)}})]+  
 3 {{{\dl}}^3} [30 {{{\dm}}^4}
\dps -\dps \ddm (10 \ddl+3
\ddm )\nonumber \\&&+ 2 {{{\dm}}^3} ({{\dps }^2}-26
\ddl-60\ddm  )- 
 2 {{{\dm}}^2} (\dps  (5 \ddl+3
\ddm )-{\lambda^{(3)}})+ 
 6 \ddm {\lambda^{(3)}}+8 \ddl {\mu^{(3)}}+4
\ddm {\mu^{(3)}}+  
 2 {\dm} (6 {{\ddl}^2}+{{\dps }^2}
(\ddl+\ddm )\nonumber \\&&+
 2\ddl  \ddm  -  
 2\dps 
({\lambda^{(3)}}+{\mu^{(3)}}))]+
 3 {{{\dl}}^2} \{33 {{{\dm}}^5}
\dps +2 {\dps ^2} \ddl \ddm +  
 17 {{\ddl}^2} \ddm +  9
\ddl {{\ddm }^2}+{{\ddm }^3} +
{{{\dm}}^4} ({{\dps }^2} -90
\ddl-165 \ddm ) \nonumber \\&& -
 2 {{{\dm}}^3} (2\dps  (\ddl +\ddm )+{\lambda^{(3)}})+  
 4 {\lambda^{(3)}} {\mu^{(3)}}- 4\dps  (\ddm 
{\lambda^{(3)}}+
\ddl {\mu^{(3)}})+
 {\dm} [-\dps  (11 {{\ddl}^2} +18
\ddl \ddm +3
{{\ddm }^2})+ 
 4 (\ddm (3
{\lambda^{(3)}}\nonumber \\&&+{\mu^{(3)}})+\ddl
  (4 {\lambda^{(3)}} +3 {\mu^{(3)}}))]+
 {{{\dm}}^2} [6 {{\ddl}^2}+{{\dps }^2}
(3 \ddl+\ddm )  -6
\ddl  \ddm - 
 2\dps  (3 {\lambda^{(3)}}+{\mu^{(3)}})]\} + 
 3 {{{\dm}}^2} \big\{3 {{\ddl}^3}+{{\dps }^2}
({{\ddl}^2}\nonumber \\&&+10 {{\ddm}^2})+  
 2{{\ddl}^2}  \ddm  -  
 4\dps  (\ddl
{\lambda^{(3)}} +10
 \ddm  {\mu^{(3)}})+2 \big(55
{{\ddm }^3}  +
 {{{\lambda^{(3)}}}^2}+10 {{{\mu^{(3)}}}^2}\big)\big\}-  
 3 {\dl} \big\{66 {{{\dm}}^5}
\ddl+9 \dps {{\ddl}^2} \ddm +
 3\dps  \ddl {{\ddm }^2}\nonumber \\&&-14
\ddl \ddm 
{\lambda^{(3)}}-  
 3 {{\ddm }^2} {\lambda^{(3)}}+{{{\dm}}^4}
  (-\dps \ddl+3 {\lambda^{(3)}})- 
 8 {{\ddl}^2} {\mu^{(3)}}-4 \ddl \ddm  {\mu^{(3)}}+  
 {{{\dm}}^2} [\dps  \ddl (9 \ddl+10 \ddm )-  
 4 (2 \ddm {\lambda^{(3)}}+\ddl (3 {\lambda^{(3)}}\nonumber \\&&+{\mu^{(3)}}))]-
 2 {{{\dm}}^3} [{{\dps }^2}
\ddl-2
{{\ddl}^2}-  6\ddl  \ddm
-2\dps  {\lambda^{(3)}}]-  
 {\dm} \big[10 {{\ddl}^3}+
 {{\dps }^2} 
 \ddl (\ddl+2 \ddm )+
18{{\ddl}^2}  \ddm-  
 4\dps  (\ddm 
{\lambda^{(3)}}+\ddl ({\lambda^{(3)}}\nonumber \\&&+{\mu^{(3)}}))+ 
 2 \big({{{\lambda^{(3)}}}^2}+2 {\lambda^{(3)}}
{\mu^{(3)}}\big)+
 2 \ddl 3 {{\ddm}^2}\big]\big\}\Big]\, .\nonumber\\
\eea
These are obtained by extracting $\lambda^{(4)}$, $\mu^{(4)}$
and $\ddot{\psi}$ terms from $h_{2}$ and $h_{3}$, respectively.

\bea
f_4 &=& 2\zeta(3)
\Big[60y_1^7y_4-36y_1^5 y_2 y_4+ 6y_{1}^6 (y_4^2-70y_2)
+22 y_1^3 y_2 (-3y_2 y_4+4y_3) \nonumber \\
& &+6y_{1}^2 (y_2^2 y_4^2+11y_2^3 -4y_2 y_3 y_4+2y_3^2)
+y_2 (y_2^2 y_4^2 +8y_2^3-6 y_2 y_3 y_4+6y_3^2)
\nonumber \\
& &+4y_1 y_2^2 (-5y_2 y_4+12y_3 )+11y_1^4
(y_2 y_{4}^2-2y_3 y_4)\Big]. \nonumber \\
\eea

\be
\begin{array}{ll}
c_{1}=9\,\zeta(3) \ap{}{}^3 \left[1680 y^{(f)}_{1}{}^7+168y^{(f)}_{1}{}^6 y^{(f)}_{4}\right],&
c_{2}=396\,\zeta(3)\ap{}{}^3y^{(f)}_{1}{}^5 y^{(f)}_{4},\\
c_{3}=-396\,\zeta(3)\ap{}{}^3y^{(f)}_{1}{}^5, & c_{4}=216\,\zeta(3)\ap{}{}^3y^{(f)}_{1}{}^7\,. 
\\
d_{1}=18\,\zeta(3) \ap{}{}^3 \left[480 y^{(f)}_{1}{}^7+ 42y^{(f)}_{1}{}^6 y^{(f)}_{4}\right], &
d_{2}=18\,\zeta(3)\ap{}{}^3 \left[6 y^{(f)}_{1}{}^6+11y^{(f)}_1{}^5y^{(f)}_4\right],\\
d_{3}=-198\,\zeta(3)\ap{}{}^3y^{(f)}_{1}{}^5,&d_{4}=108\,\zeta(3)\ap{}{}^3y^{(f)}_{1}{}^7\,. \\
e_{1}=2\,\zeta(3) \ap{}{}^3 \left[420 y^{(f)}_{1}{}^6y^{(f)}_{4}+ 36y^{(f)}_{1}{}^5 y^{(f)}_{4}{}^2\right],&
e_{2}=2\,\zeta(3)\ap{}{}^3 \left[11 y^{(f)}_{1}{}^4y^{(f)}_{4}{}^2 -36y^{(f)}_1{}^5y^{(f)}_4-420y^{(f)}_1{}^6\right], \\
e_{3}=-44\;\zeta(3)\ap{}{}^3y^{(f)}_{1}{}^4y^{(f)}_{4},&
e_{4}=2\,\zeta(3)\ap{}{}^3\left[60y^{(f)}_{1}{}^7+12y^{(f)}_1{}^6y^{(f)}_4\right].
\end{array}
\ee
\bea
s_{1}&=& y_{4}-e_{1}+36\zeta(3)\ap{}^3 y_{1}^5 (y_4^2+9y_1^2-270\zeta(3)\ap{}^3 y_1^8) +6\zeta(3)\ap{}^3 y_{1}^6 (18y_1+2d_1-c_1)\\ \nonumber
&&-9wy_1+\frac{w}{2} c_1\\
s_{2}&=& -1-e_{2}+11\zeta(3)\ap{}^3 y_{1}^4 (y_4^2+9y_1^2-270\zeta(3)\ap{}^3 y_1^8) +6\zeta(3) \ap{}^3 y_1^6 (2d_2-c_2)+\frac{w}{2}c_{2} \\
s_{3}&=& -e_{3}+6\zeta(3)\ap{}^3 y_1^6
(2d_{3}-c_{3})+\frac{w}{2}c_{3} \\
\label{s4}
s_{4}&=& y_{1}-e_{4}+6\zeta(3)\ap{}^3 y_1^6
(2d_{4}-c_{4}+2y_4)+wy_4+\frac{w}{2} c_{4}
\eea

\newpage
%%%%%%%%%%%%%%%%%%

\end{document}